\documentclass{raa}

\usepackage{graphicx,times}             
\usepackage{natbib,amssymb}
\bibpunct[, ]{(}{)}{;}{a}{}{,}

\def\beq{\begin{equation}}
\def\eeq{\end{equation}}
\def\bear{\begin{eqnarray}}
\def\ear{\end{eqnarray}}

\begin{document}

   \title{Galactic Center Research: Manifestations of the Central Black Hole}
   

   \volnopage{Vol.0 (200x) No.0, 000--000}      
   \setcounter{page}{1}          

   \author{Mark R. Morris
      \inst{1}
   \and Leo Meyer
      \inst{1}
   \and Andrea M. Ghez
      \inst{1}
   }

   \institute{Department of Physics and Astronomy, University of California, Los Angeles, CA, USA 90095-1547; {\it morris@astro.ucla.edu}\\
   }


\abstract{This review summarizes a few of the frontiers of Galactic center research that are currently the focus
of considerable activity and attention.  It is aimed at providing a necessarily incomplete sketch of some of the timely work being done on phenomena taking place in, or originating in, the central few parsecs of the Galaxy, with particular attention to topics related to the Galactic black hole (GBH).  We have chosen to expand on the following exciting topics: 1) the characterization and the implications for the variability of emission from the GBH, 2) the strong evidence for a powerful X-ray flare in the Galactic center within the past few hundred years, and the likelihood that the GBH is implicated in that event,  3) the prospects for detecting the "shadow" of the GBH, 4) an overview of the current state of research on the central S-star cluster, and what has been learned from the stellar orbits within that cluster, and 5) the current hypotheses for the origin of the G2 dust cloud that is projected to make a close passage by the GBH in 2013.
\keywords{Galaxy: center --- black hole physics --- X-rays --- infrared radiation --- ISM: clouds}}

   \authorrunning{M. R. Morris, L. Meyer \& A. M. Ghez }            
   \titlerunning{Recent Results on the Galactic Black Hole }  

   \maketitle

%
%
\section{Introduction}           
\label{sect:intro}

If we adopt the discovery of the nuclear stellar cluster in the infrared by \citet{BN68} as the beginning of research focussed on the central parsecs of the Galactic center, the field is only 44 years old.   In that time, the pace of discovery has been breathtaking, and it has not shown any signs of leveling off.  This review is intended to illustrate this fact by offering brief descriptions of some selected samples of exciting research trends that are currently under intense investigation.  It focusses on topics that are related in various ways to the supermassive Galactic black hole, often given the name of the associated radio source, Sagittarius A*.  The review is not meant to be complete; many important topics are left out.  For comprehensive overviews on topics not addressed here, the reader is referred to the recent review article by \citet{GEG10} and to the proceedings of a recent international conference on this subject \citep{MWY11}.   

\section{The time variability of Sgr~A* in the X-rays, near-infrared, and radio}

In this section, the discussion is largely restricted to the observational state of affairs for the variability of Sgr A*.  There exists a substantial literature on the theoretical interpretation of the emission from Sgr A* and its variability that we only briefly address here.  For access to the rich variety of theoretical models, the reader is referred to the references just mentioned, to the review by \citet{MeliaFalcke}, or to the manifold references in the observational papers that we do cite, such as those in \citet{marrone08}, \citet{eckart12}, or \citet{Yusef-Zadeh12}.  

\subsection{The near-infrared properties of Sgr~A*}
\label{subsect:NIR} 

\begin{figure}
    \centering
        \includegraphics[width=13cm]{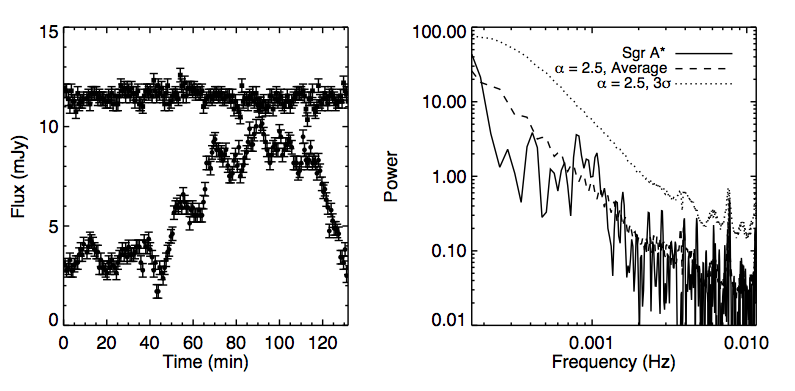}

        \vspace{-3mm}
        \caption{Left panel: The extinction-corrected light curve of Sgr~A*, from 2003 June 16, as extracted by \citet[][lower curve]{meyer06b} along with a comparison source (upper curve, offset by 6 mJy for clarity). This data set was first reported by \cite{genzel03}. Right panel: The periodogram of the Sgr~A* light curve (solid) with the mean power (dashed) and 3$\sigma$ thresholds set by Monte Carlo simulations of red noise with a power-law slope of $\alpha = 2.5$ (dotted). There is no significant peak in the periodogram and therefore no evidence for a quasi-periodicity. From \cite{do09}.}
        \label{flare}
\end{figure}

The near-infrared (NIR) flux associated with the accretion flow around Sgr~A* was independently discovered by the UCLA and MPE groups in 2002, when adaptive optics instruments became available. Since the bright short-period star S0-2 was going through closest approach at that time, the two sources were blended and the measurements were somewhat ambiguous. A year later, S0-2 was sufficiently far away and a variable red source at the dynamical center of the galaxy could unambiguously be identified \citep[][see also Figure~\ref{flare}]{genzel03, ghez04}.

Ever since its discovery it was claimed that Sgr~A*'s variability is accompanied by a quasi-periodicity (QPO) of $\sim$20 minutes. However, better analysis techniques and more data could not substantiate that claim (Figure~\ref{flare}). It is now accepted that there is no statistical evidence for a persistent, stable QPO in the frequency region that can be probed \citep{meyer08, do09, dodds11}. While there is no clear evidence for transient quasi-periodic phenomena, it has been shown that polarimetric observations of Sgr~A*'s flux excursions are consistent with a "hot spot" model in which the emission from a bright region of the accretion flow is enhanced by strong gravitational lensing and Doppler boosting \citep{eckart06, meyer06a, meyer06b, meyer07, trippe07, zamani10}. 

\begin{figure}
    \centering
        \includegraphics[width=10cm]{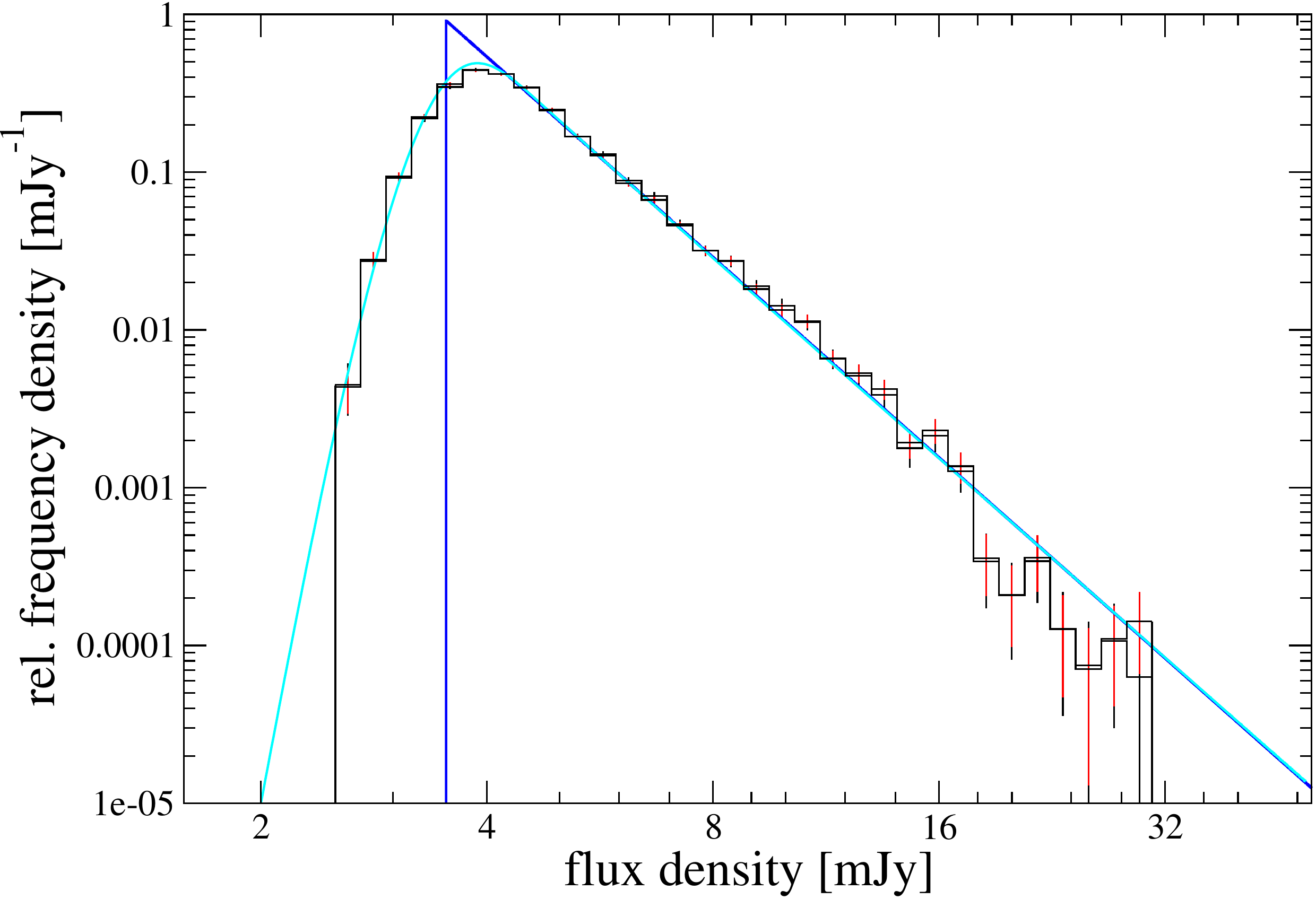}

        \vspace{-3mm}
        \caption{The double log histogram of the (confusion  and extinction corrected) flux density of Sgr~A* as obtained from all Ks-band observations with VLT/NaCo between 2003 and summer 2010. The observed data - 10639 data points - are depicted in black with black error-bars, the red error bars display the uncertainty with respect to randomly drawn subsamples of the data, the blue line represents the best intrinsic power-law model of the variable process, and the cyan line is the best fit to the data, obtained by convolution of the best power-law model with a Gaussian distribution to account for the instrumental error. The break of the power-law at small flux densities corresponds to an intrinsic source flux of zero. There is no need to assume distinct states of Sgr~A*. From \cite{witzel12}.}
        \label{hist}
\end{figure}

The intrinsic NIR variability of Sgr~A* can be modeled as a purely random process.  \cite{meyer08} show this for the longest continuous light curve observed so far (600 minutes), while others have analyzed all available data sets from the Keck Observatory \citep{do09} and the VLT \citep{dodds11,witzel12}.  The latter two papers analyze around $\sim10^5$ flux measurements taken in $\sim$7 years.  The distribution of  measured fluxes is shown in figure~\ref{hist}. The interpretation of the flux distribution offered by \cite{dodds11} and \cite{witzel12} is quite different: \cite{dodds11} use a log-normal + power-law distribution (convolved with a Gaussian to account for measurement errors) to describe the distribution and argue that Sgr~A* has two distinct states, one described by the log-normal part, the other by the power-law tail. In contrast, \cite{witzel12} find that only a power-law (convolved with a Gaussian) is needed to accurately describe the intrinsic flux distribution of Sgr~A*.

\begin{figure}
    \centering
        \includegraphics[width=10cm]{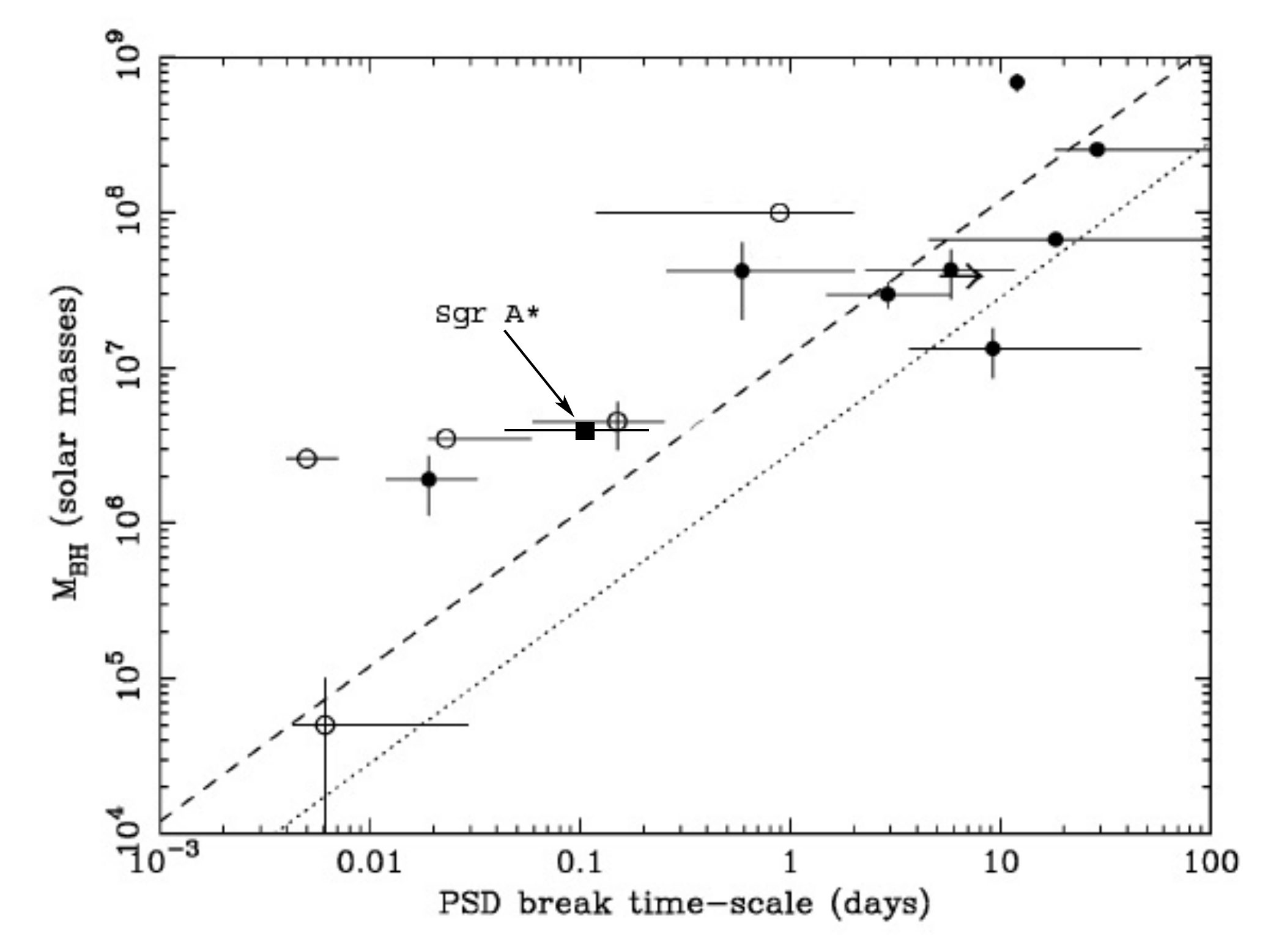}

        \vspace{-3mm}
        \caption{Sgr~A*'s break timescale as reported by \cite{meyer09} over plotted onto Figure~11 of \citet{uttley05}, which shows BH mass versus PSD break timescale for various AGN. The mass of Sgr~A* has been taken from \citet{ghez08}. Its uncertainty corresponds to the height of the black square. Filled circles mark masses determined from optical reverberation mapping, open circles represent masses determined using other methods. The straight lines represent the expected relations if linear mass scaling is assumed from the typical timescales observed in the high/soft (dashed line) and low/hard (dotted line) state of the BHXRB Cyg X-1 (assuming 10 solar masses for its mass). From \cite{meyer09}.}
        \label{break}
\end{figure}

The power-spectral density (PSD) of Sgr~A*'s time variability is a featureless power-law for relatively high frequencies (see figure~\ref{flare}). \cite{meyer09} find that the power-law breaks to a shallower slope at lower frequencies; the break corresponds to a time of $154^{+124}_{-87}$\,minutes (errors mark the 90\% confidence limits). Such a break of the PSD power-law is also found in other black hole accretion systems, enabling  a comparison of this characteristic timescale across a wide range of masses (figure~\ref{break}). \cite{meyer09} find that the timescale in Sgr~A* is inconsistent with a recently proposed scaling relation that uses bolometric luminosity and black hole mass as parameters \citep{mchardy}. However, the result fits the expected value if only linear scaling with black hole mass is assumed. The authors therefore suggest that the luminosity-mass-timescale relation  applies only to black hole systems in the soft state. In the hard state, which is characterized by lower luminosities and accretion rates, there seems to be just linear mass scaling, linking Sgr~A* to hard-state stellar mass black holes. 

\subsection{X-ray variability of SgrA*}

A relatively weak X-ray counterpart to Sgr A* was discovered in 1999, early in the initial operations of the Chandra X-ray Observatory \citep{Baganoff03}.  In the years since then, the quiescent emission level has changed very little; it is likely to arise relatively far from the black hole, at the Bondi accretion radius where the accretion flow undergoes shocks and thermalizes, producing bremsstrahlung X-rays \citep{Quataert02}.  With Chandra, this source is marginally resolved \citep{SB10}, with an intrinsic width of about 0.6", which corresponds well with the expected size of the Bondi radius.  Another mechanism that may contribute to (or possibly even dominate) the extended emission is coronal radiation from a population of spun-up low-mass stars \citep{SSR12}.  

Soon after the initial discovery of X-rays from Sgr A*, a large X-ray outburst lasting almost 3 hours and having a peak intensity of 36 times the quiescent level was reported by \citet{Baganoff01}.  Numerous X-ray flares have been observed since then by both Chandra and XMM-Newton.  Their variability time scales, including temporal structure on several-minute time scales during the flares, indicate that they arise much closer to the black hole than the quiescent emission, probably within a few tens of Schwarzchild radii.  The first X-ray flare reported was stronger and longer in duration than most flares, which are typically $\sim$30 minutes in length and have peak fluxes that range between a few and ten times the quiescent flux.  The recurrence rate of detectable flares is about once per day.  However, this may be limited by sensitivity; weaker flares probably go undetected and may contribute to the mean inferred quiescent flux.  Since flares are point sources, the undetected ones could also lead to a narrowing of the perceived size of the quiescent emission region.  

\begin{figure}
    \centering
        \includegraphics[width=10cm]{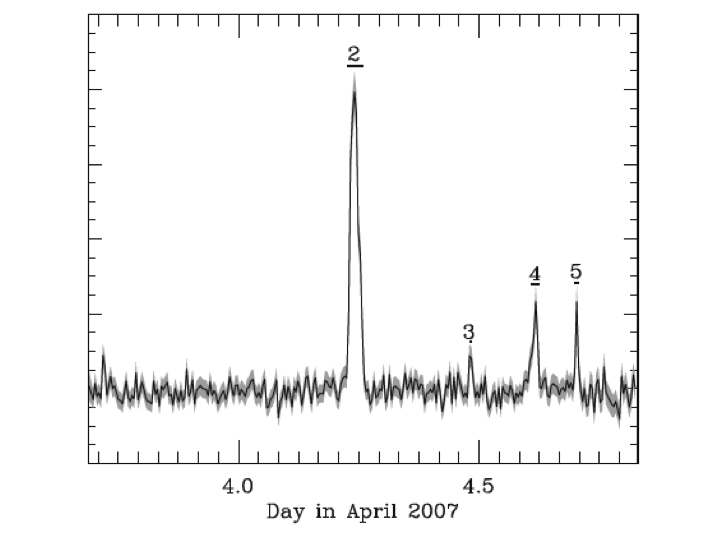}

        \vspace{-3mm}
        \caption{Light curve of one of the three brightest flares that have been found so far. From 
        \citet{Porquet08}, using XMM-Newton.  The flares following the main flare are presumably 
        "aftershocks" related to the main event.} 
        \label{porquet08flare}
\end{figure}

There have been three particularly bright flares detected from Sgr A*, two by XMM \citep{Porquet03,Porquet08} and one recently by Chandra/HETG \citep{Nowak12}, all having peak fluxes on the order of 100 times the quiescent level.  The latter flare, lasting two hours, is the first result to be reported from a Chandra Cycle 13 {\it X-ray Visionary Project} to study Sgr A*, in which a total of 3.2 Msec is being dedicated to observations with the High-Energy Transmission Grating Spectrometer.  The three bright flares all have spectral slopes, $\Gamma$, of around 2.  \citet{Nowak12} suggest that it is possible that this slope is typical of all of the flares, although various instrumental issues have prevented accurate spectral slope determinations for weaker flares.  The bright flares, including the first flare reported by Baganoff, also typically display small precursor or aftershock flares (see, e.g., figure \ref{porquet08flare}), and this behavior may also be typical of all flares, although it is generally not detected for weaker flares because of inadequate sensitivity.  

When simultaneous X and IR observations can be carried out, as has been done many times, it appears that every X-ray flare corresponds reasonably well in time with a well-defined and relatively bright maximum in the infrared light curve \citep{eckart04,ghez04,marrone08,dodds-eden09,trap11,eckart12}.\footnote{The near-infrared maxima are quite often also referred to as "flares," although as the discussion above indicates, they are more aptly referred to as peak excursions in a continuously and randomly varying light curve.  This raises the question of whether the X-ray peaks are also part of a continuously variable light curve that is, however, much more extreme in its variability amplitude so that it is well below the detection threshold most of the time}  The reverse is not true: only a fraction of the IR maxima correspond with an X-ray flare.    
The X-ray emission is universally considered to be nonthermal, and there has been some debate about whether the X-ray flares are the result of direct synchrotron emission, inverse Compton emission, or is the result of a synchrotron self-Compton process, wherein the sub-mm or infrared synchrotron photons emitted by a population of relativistic electrons are inverse Compton scattered up to X-ray energies by the same population of electrons.  The current preponderance of opinion favors the latter, and the reader is referred to the papers referenced here for detailed discussions of the various alternatives.  A recent report by \citet{Yusef-Zadeh12} of a significant time lag between the infrared maxima and the X-ray flares might elucidate this issue, and help determine from where in the accretion flow the emission is emerging.  

\subsection{Variability of sub-millimeter and radio emission}

The radio variability of Sgr A* has been studied for over three decades \citep{BrownLo82}, although the centimeter-wavelength variability initially observed is largely attributable to interstellar scintillation, rather than to intrinsic variability \citep{Zhao89,MacBow06}.  The scintillation timescales are wavelength-dependent, giving rise to trends lasting from weeks to years.  At shorter wavelengths, the scintillation amplitude declines, and monitoring campaigns have been undertaken to tease out the intrinsic variability in the light curves, with indications that on time scales of hours to days, flaring events are occurring \citep{bower02,Zhao03,Herrnstein04,Miyazaki04}. The intrinsic radio variability of Sgr A* is best seen on such time scales.  In an interferometric study at $\lambda$3mm, \citet{Mauerhan05} characterized the variability as a red noise process with amplitudes up to 40\%.  And in observations at 880 $\mu$m with the Submillimeter Array, \citet{marrone06} reported strong variability of both total intensity and polarization fraction on a time scale of hours.  Interestingly, the power spectrum reported by \citep{Mauerhan05} showed a suggestive maximum on a time scale of $\sim$2.5 hours, which coincides with the break in the power-law slope of the infrared power spectrum ({\it c.f.}, section \ref{subsect:NIR}).  

Strong clues to the nature of the emitting region can be obtained by comparing the light curves of Sgr~A* at different wavelengths.  \citet{Yusef06} carried out a VLA study in which they rapidly switched between frequencies of 22 and 43 GHz, and obtained two effectively simultaneous light curves covering an emission maximum, or flare.   Their data show a delay of the 22 GHz light curve, relative to that at 43 GHz, by 20 to 40 minutes.  They interpreted this in terms of the expanding plasmon model of \citet{vdL66} in which the plasma cools and becomes optically thin as it expands, and thus reaches an emission maximum, at progressively longer wavelengths with time.  In a subsequent study, \citet{Yusef08} coupled the same two frequencies with a simultaneous (submm) 350 GHz observation and a Chandra observation showing an X-ray flare, and found that the X-ray flare precedes the submm flare by about 90 minutes, as the expanding plasmon model would qualitatively predict, and again that the 22 GHz emission lags the 43 GHz emission.  (Curiously, however, the submillimeter maximum followed the 22 and 43 GHz maxima, which is not consistent with the expected trend.)

\begin{figure}
    \centering
        \includegraphics[width=9cm]{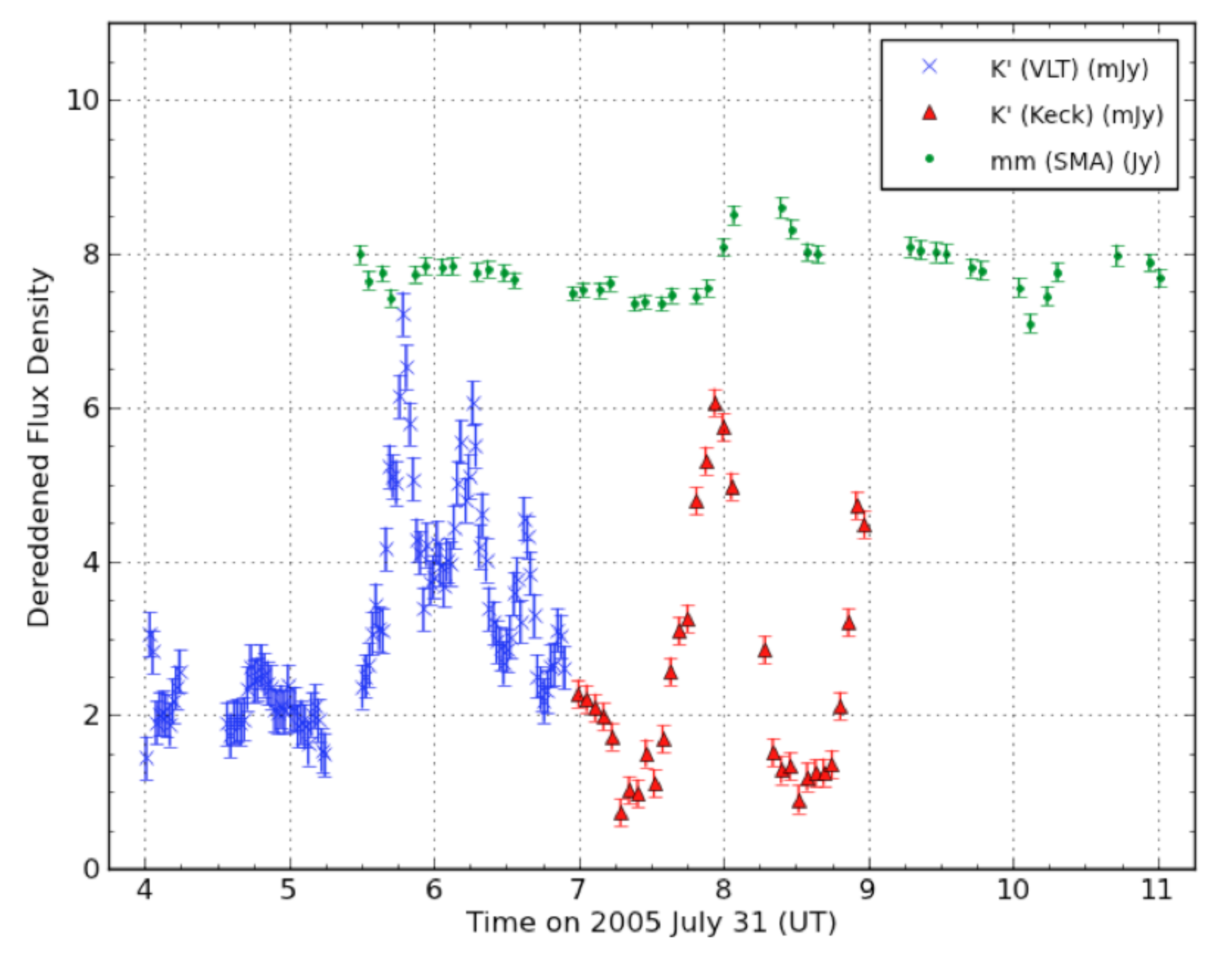}  

        \vspace{-3mm}
        \caption{Light curves of Sgr A* on 2005 July 31 in the near-infrared (lower sets of points) and 1.3-mm continuum \citep[upper green points, from the SMA;][]{marrone08}.  Blue points are from the VLT, and red points from the Keck Observatory.  Flux units are mJy for the NIR and Jy for the mm data.  The mm data have been shifted upwards by 3.5 Jy for clarity. Taking only the Keck data into account, \citet{marrone08} derived a time lag of 20 min between the mm- and the NIR-lightcurves. When using the whole data set, the cross-correlation function peaks at a time lag of 160 min. }
        \label{mm_delay}
   \end{figure}

There have, in fact, been a number of claims for a delay of a few hours between near-infrared maxima and  subsequent, associated millimeter/submillimeter peaks.   Table \ref{Tab:delays} lists the reports to date.  Individually, the reported delays are subject to question because of the uncertainties of the light curves and the non-zero possibility of incorrect matchings, but collectively, the delays all appear in the neighborhood of about 150 minutes, with one exception.  That exception is shown in Figure \ref{mm_delay}.  Originally, only the light curve from the Keck Telescope (red points) was considered by \citet{marrone08}, and there appeared to be a rather brief, 20-minute delay between the near-infrared and 1.3-mm peaks.  However, when the VLT observations (blue points) are included in the light curve, a second candidate near-IR maximum for the counterpart to the mm peak becomes evident, with a 160-minute time lag \citep{meyer08}.  A time delay of this order is consistent with an expanding relativistic plasma blob model if the expansion occurs at $\sim$0.1c \citep{marrone08,Yusef09}.  

\begin{table}
\begin{center}
\caption[]{Reported time lags between near-infrared and millimeter/submillimeter peaks in the SgrA* light curve.  References: [1] \citep{eckart06a}, [2] \citep{eckart09}, [3] \citep{Yusef06a}, [4] \citep{marrone08}, [5] \citep{Yusef08}, [6] \citep{Yusef09}, [7] \citep{eckart08}, [8] \citep{trap11}, [9] this work.  
}\label{Tab:delays}


 \begin{tabular}{clcl}
  \hline\noalign{\smallskip}
References &  Date      & Wavelength ($\mu$m) & Time Lag (min)                    \\
  \hline\noalign{\smallskip}
1,2  & 2004 Jul. 7     &  890  & $<$ 120               \\ 
3     & 2004 Sep. 3   &   850  & 160                      \\
4,5,6   & 2006 Jul. 17     &   1300, 850 & 100       \\
6     & 2007 Apr. 5   &   1300   &  160                   \\
7     & 2008 Jun. 3   &   870  &   90                       \\
8     & 2009 Apr. 1   &   870  &  200                      \\
4     & 2005 Jul. 31   &   1300  & 20                      \\
9     & 2005 Jul. 31   &   1300  & 160                     \\

  \noalign{\smallskip}\hline
\end{tabular}
\end{center}
\end{table}

The expanding blob model would predict that X-ray and near-IR emission peaks should occur simultaneously, since they are both optically thin throughout the expansion.  This is consistent with almost all observations that have been made of X-ray flares during which infrared observations were carried out.  However, \citet{Yusef-Zadeh12} have reanalyzed a body of existing data, and have concluded that the X-ray flare emission {\em lags} the near-IR maxima by a few to tens of minutes.  They therefore argue that, while the near-IR events occur in the inner portions of the accretion flow, the X-ray flares result from inverse Compton scattering of the IR photons by thermal electrons (kT$_e$ $\sim$ 10 keV) in the accretion flow, possibly at a greater radius, which occasions the delay \citep{Yusef09,wardle11}.


\section{X-Ray fluorescence in the Galactic center: echo of a recent accretion event?}
\label{sect:Fluor}

\begin{figure}
    \centering
        \includegraphics[width=9cm]{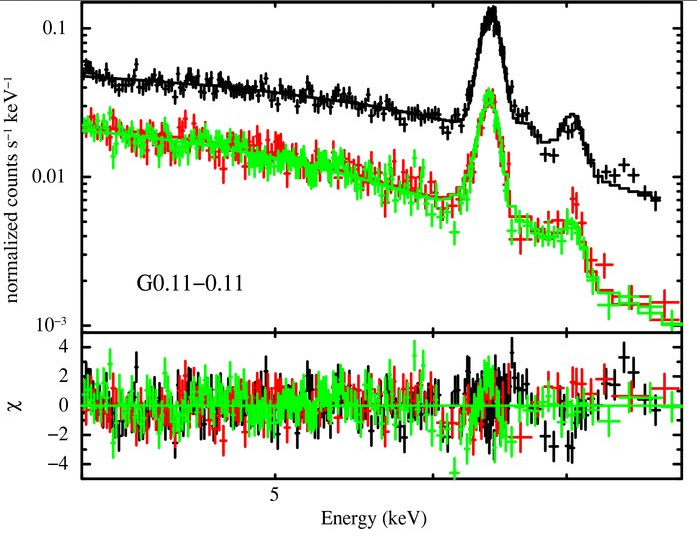}  

        \vspace{-3mm}
        \caption{X-ray spectrum of G0.11-011, one of the brighter clouds displaying emission from the neutral 
        K$\alpha$ line. One can see the Fe K$\alpha$ (6.4 keV), K$\beta$ (7.06 keV), the K edge (7.1 keV) 
        and the flat power-law continuum.   The three colors show data from the three EPIC cameras on 
        XMM-Newton.  From \citet{Ponti10}.}
       
        \label{G0.11}
   \end{figure}

It is becoming increasingly evident that one or more powerful pulses of relatively hard X-rays is propagating across the central molecular zone of the Galaxy, perhaps originating in the recent past from the central black hole, and giving rise to a "light echo" consisting of a combination of scattered hard X-rays and fluorescent X-ray line emission.  The most prominent line is the 6.4 keV K$\alpha$ line of neutral or weakly ionized iron, which results when the most tightly bound electron is ejected by an X-ray with an energy above the 7.1 keV ionization edge of iron, and in the ensuing radiative cascade, K lines are emitted, with K$\alpha$ being the most intense.  
This line arises from numerous molecular clouds near the center (e.g., Figure \ref{G0.11}), and the emerging interpretation is that these are the clouds that are currently being illuminated by an external, variable or transient source.   \footnote{Strong K$\alpha$ emission is also observed throughout the Galactic center region from highly ionized, helium-like iron at 6.7 keV, and that is attributed to a $\sim$10$^8$ K thermal source consisting of some combination of: (1) supernova remnants, (2) an unresolved collection of X-ray emitting binary stars, and (3) a coronal gas extending over several hundred parsecs around the Galactic center.  The relative contributions of (2) and (3) are presently under debate; see the review by \citet{Goldwurm11review}.}  

Fluorescent production of the 6.4 keV line was first suggested by \citet{SMP93}.  These authors used the broadband flux measured by the ART-P telescope on the GRANAT satellite to argue that the luminosity of the accretion flow onto the Galactic Black Hole (Sgr A*) cannot have been much higher in the past several hundred years than it is at present.  Subsequently, \citet{Koyama96} performed imaging and spectra with the ASCA satellite and produced the first map of 6.4 keV line emission.  The strongest of two prominent emission regions was centered on the massive molecular cloud, Sgr B2, while the other was located in the vicinity of the Galactic center Radio Arc.   \citet{Koyama96} proposed that an energetic flaring event of Sgr A* and occurring about 300 years ago -- the light travel time between the Sgr A* and Sgr B2 -- is responsible for the fluorescent emission, and they noted that a flare luminosity as high as 2 $\times$ 10$^{39}$ ergs s$^{-1}$ would be required to account for the present-day fluorescent emission.  

An alternative to X-ray-induced fluorescence is collisional ionization by low-energy cosmic ray electrons, typically $\sim$30 keV).  This model was proposed by \citet{Valinia2000} to account for the X-ray spectrum and the 6.4 keV line emission from the large-scale Galactic ridge, and was adopted by \citet{YLW02} to explain the unusual brightness of the 6.4 keV line in the G0.11-0.11 (also known as G0.13-0.13) molecular cloud in the Galactic center (Park et al. 2004), which is apparently interacting with relativistic particles in the Galactic center Radio Arc.  \citet{YLW02} suggested that the same mechanism might be applicable to 6.4 keV line sources elsewhere in the Galactic center.  Subsequently, \citet{Yusef-Zadeh07a} explored this hypothesis in detail by producing a new map of 6.4 keV line emission based on observations with Chandra, and arguing that there exists a correlation between nonthermal radio continuum emission from magnetized filaments and the 6.4 keV line emission and hard X-ray continuum.  They account for this correlation in terms of the impact of relativistic particles from the nonthermal filaments upon neutral gas, producing both nonthermal bremsstrahlung X-ray continuum emission and diffuse 6.4 keV line emission.  In addition, they propose other consequences of this interaction, including cosmic-ray heating of molecular gas in the Galaxy's central molecular zone and the production there of large-scale diffuse TeV emission by Compton up-scattering of thermal sub-mm photons from dust.  A related hypothesis -- K-shell electron ejection by the impact of low-energy cosmic ray {\it protons} emitted by Sgr A* was suggested by Dogiel et al. (2009), but \citep{Chernyshov12} argue that this mechanism would require rather extreme parameters, so is a less likely model for production of diffuse 6.4 keV line emission than K shell vacancy production by X-rays.

\begin{figure}
    \centering
        \includegraphics[width=12cm]{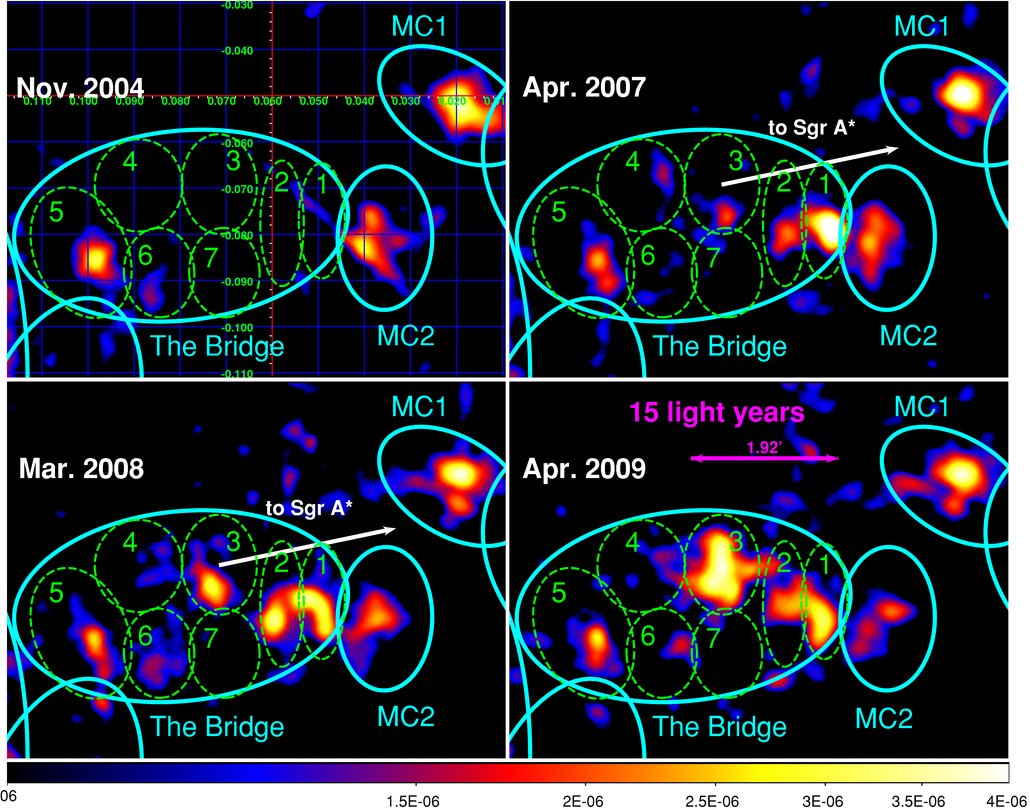}  

        \vspace{-3mm}
        \caption{Fe K$\alpha$ continuum-subtracted mosaic image of the observations of the "bridge" region made with the EPIC-pn camera on XMM.  The ellipses outline the various regions for which light curves were obtained by \citet{Ponti10}.  Variations are evident on time-scales of about 2-4 years, but on spatial scales of about 15 light years. The implied superluminal motion can be attributed to illumination by a bright (L $>$ 1.3 $\times$ 10$^{38}$ erg s$^{-1}$) and distant ($>$15 pc) X-ray source active for several years. From \citet{Ponti10}.}
       
        \label{Kalpha_maps}
   \end{figure}

A number of results since 2007 have rather strongly tipped the balance in favor of the X-ray fluorescence hypothesis.  First, Muno et al. (2007) found rapid changes in the intensities and morphologies of two relatively compact, hard X-ray continuum nebulosities, a result that could be explained by a moving reflection nebula created by an X-ray front propagating at lightspeed, but which is not consistent with the rates of diffusion expected for low-energy cosmic rays.  This result was greatly strengthened by Ponti et al. (2010) who used $\sim$1.2 Msec of X-ray data from the XMM-Newton satellite spanning about 8 years.  They reported a complex pattern of variations of 6.4-keV line emission in numerous molecular clouds within 15 arcminutes ($\sim$40 pc projected) of Sgr A* (figure \ref{Kalpha_maps}), with the emission from some clouds increasing, some remaining constant, and some decreasing \citep[see also][]{Capelli12}, including the bright source G0.11-0.11 that had been invoked as an illustrative example for the low-energy cosmic ray hypothesis.  Ponti et al. interpret their overall results in terms of apparent superluminal motion of a light front propagating through the region, and they find that all of the observations are consistent with a single flare from Sgr A* or its vicinity if the luminosity of that flare is at least $\sim$1.5 $\times$ 10$^{39}$ ergs s$^{-1}$ and if it faded about 100 years ago.  Potential sources other than Sgr A* could not be ruled out, but the required luminosity is at or above that which could be produced by a binary system.  The apparent superluminal proper motion of the fluorescent emission had been anticipated by \citet{SunyaevChurazov98}.  

\subsection{Sgr B2}
Originally the most prominent Galactic center source of Fe K$\alpha$ emission, the massive Sgr B2 molecular cloud is a key object that has recently undergone a particularly dramatic transition.  \citet{Murakami01b} were the first to report Chandra observations of SgrB2, and they found that the diffuse X-rays, showing prominent K$\alpha$ and K$\beta$ peaks from neutral iron, were distributed with a concave morphology wrapping around Sgr B2, and arise mainly from the side of the cloud facing SgrA*.  Their detailed model for the distribution of 6.4 keV line emission matched the observations well, and supported the original X-ray fluorescence model for this source based on ASCA observations \citep{Koyama96}.  No  source local to Sgr B2 had sufficient X-ray continuum flux to account for the 6.4 keV emission via fluorescence.  

The Suzaku satellite observed Sgr B2 in 2005, 5 years after the Chandra observation, and \citet{Koyama08} found that its 6.4 keV flux had dropped by a factor of two in that interval.  This decline is evident in a 10-year "light curve" from 1995 to 2005 assembled by \citet{Inui09} from the six available measurements at that time, and another 2009 Suzaki point shows the decline continuing to that point \citet{Nobukawa11}.  Because the light-crossing time of the central parts of the cloud is comparable to this time scale for the decline, \citet{Koyama08} attribute the decrease to the the passage of the tail end of the flare from Sgr A*.  Again, a Sgr A* luminosity of at least 2 $\times$ 10$^{39}$ ergs s$^{-1}$ would be required before the decline.  

\begin{figure}
    \centering
        \includegraphics[width=10cm]{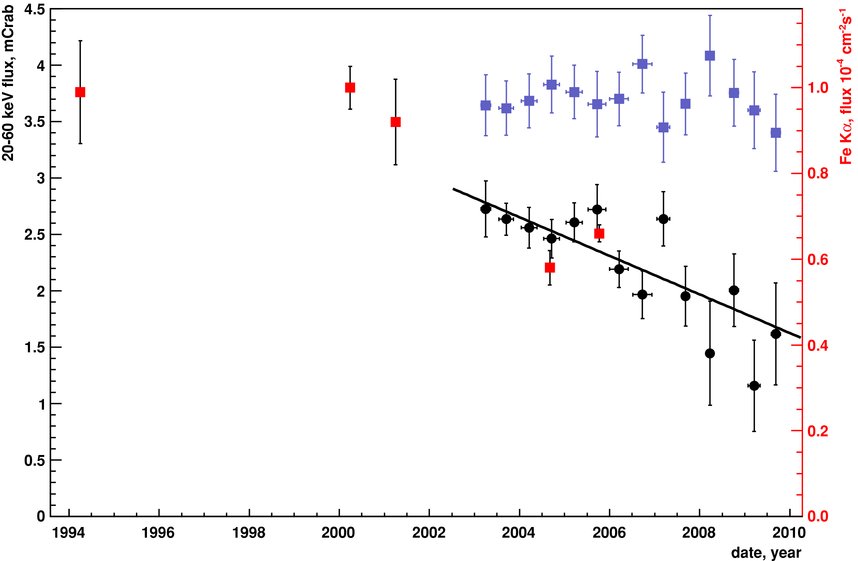} 

        \vspace{-3mm}
        \caption{Light curve of the corrected 20 - 60 keV flux of Sgr B2 as measured with IBIS/ISGRI on the INTEGRAL satellite from 2003 to 2009 (Black circles). The light curve of the secondary calibrator (Ophiuchus cluster) is shown for comparison (light blue squares). Superimposed red squares show the Fe K$\alpha$ line fluxes obtained by Inui et al. (2009) with ASCA, Chandra, XMM-Newton, and Suzaku observations taken from 1994 to 2006. The hard X-ray flux shows a clear decay
of up to 40\% of the measured Sgr B2 flux from 2003 to 2009, and the 6.4 keV measurements show a similar trend after the year 2000, with the flux being apparently constant from 1994 to 2000. From \citet{Terrier10}.}
       
        \label{fade}
   \end{figure}

The fluorescent emission should be accompanied by Compton scattering of the hard X-rays impinging on the cloud. Indeed, data from the INTEGRAL satellite show that Sgr B2 coincides with the diffuse hard X-ray source IGR J17475-2822, which has substantial flux (2.5 $\pm$ 0.1 mCrab) even in the 20 - 200 keV band \citep{Revnivtsev04}.  A more recent study based on the complete set of INTEGRAL observations from 2003 to 2009 \citep{Terrier10} shows that the decay time of the hard X-ray flux from this source is 8.2 $\pm$ 1.7 yr, and this closely accompanies the decline of neutral iron K$\alpha$ intensity, as shown in Figure \ref{fade}.  The SUZAKU observations of \citep{Nobukawa11} confirm this decline of the diffuse, hard X-ray flux.  \citep{Terrier10} reexamine the possibility that low-energy cosmic ray electrons are responsible for the observed X-ray characteristics, and find that an injection power of order 10$^{40}$ ergs s$^{-1}$ would be needed.  That and the short decay time lead them to rule out that scenario. 

In principle, the details of the morphology of scattered and fluorescing X-rays, its spectrum, and the time variations of both can all be used as diagnostics of the structure of a cloud, the placement of the cloud with respect to the source of the incident flare, and the temporal profile of that flare \citep{SunyaevChurazov98,Odaka11}.  However, this is clearly a complex inverse problem.  In their theoretical study motivated by Sgr B2, \citet{Odaka11} point out that, because X-rays with energies above 20 keV can penetrate much deeper into a cloud, the distribution of the ratio of the hard continuum emission to the intensity of the 6.4 keV fluorescent line can be a useful additional probe for the internal cloud structure, as well as for its relative placement.  The fading of Sgr B2 is will unfortunately complicate the future application of these models to that one cloud, but the expanding spherical shell of X-rays that underlies the current paradigm should eventually illuminate another cloud in the central molecular zone.  

\subsection{Sgr C}
\begin{figure}
    \centering
        \includegraphics[width=11cm]{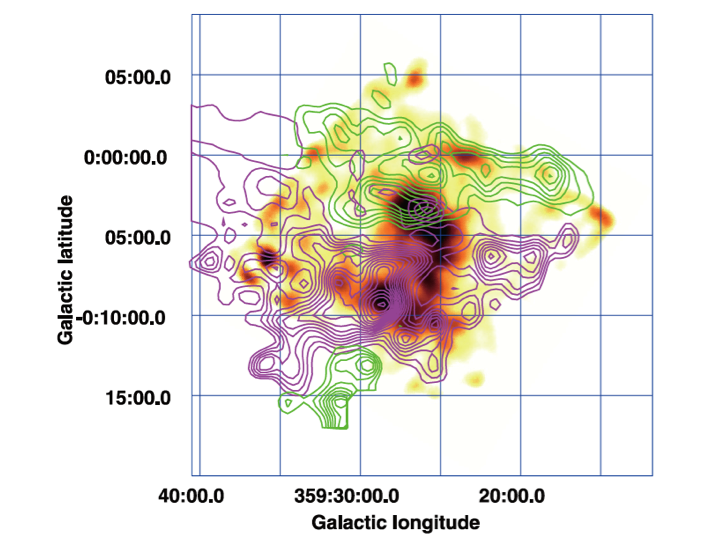}

        \vspace{-3mm}
        \caption{Contours of two channel maps of CS J = 1-0 line emission overlaid on the Suzaku image of 6.4 keV neutral iron line emission, from \citet{Nakajima09}.  Green (magenta) contours show emission in the velocity range -90 to -130 km s$^{-1}$ (-30 to -80 km s$^{-1}$).  The CS data are from \citet{Tsuboi99}.}
       
        \label{SgrC}
   \end{figure}

Opposite the Galactic center from Sgr B2 (Galactic longitude, l = 0.7$^{\circ}$) lies another well-studied massive molecular cloud, Sgr C (at l = -0.5$^{\circ}$).  Using ASCA observations, \citet{Murakami01a} identified it as another example of an X-ray reflection nebula because of its diffuse, hard X-ray emission and strong 6.4 keV line.  As with Sgr B2, there was no adequately bright X-ray continuum source in the immediate vicinity of Sgr C to account for the observed fluorescence.  The correspondence between the molecular gas and the 6.4 keV line emission in SgrC is well displayed in figure \ref{SgrC}, from \citet{Nakajima09}.  Comparing earlier data from ASCA and Chandra with their Suzaku measurements, \citet{Nakajima09} raise the possibility that there has been variability in the 6.4 keV emission among the several source components, but   
because of the different spatial resolutions and sensitivities of these three instruments, the comparison is not conclusive.  

\subsection{Is Sagittarius A* the source of the X-ray flash?}
Since the first discussion of the fluorescence phenomenon by \citep{SMP93}, Sgr A* has been the primary suspect for producing the bright X-ray flash.  The phenomenon is apparently distributed throughout the central molecular zone, so if a single source is responsible, its high luminosity requirement probably places it above the range of luminosities that have been exhibited by X-ray binaries.   The presence of numerous distributed sources with X-ray binary characteristics has been considered as an alternative hypothesis, but none with sufficient intensity has been found in the vicinity of the fluorescing clouds, so that hypothesis is not being pursued.  

If a single source is responsible for the moving fluorescent wall, observations should eventually provide a series of intersecting vectors that all point back to the source.  That is not available yet because the apparent progression of the X-ray front is complicated by the unknown internal structures of clouds.  The superluminal movement of the front along an obliquely oriented density ridge, for example, can give the impression that the source is aligned with the ridge.  Nonetheless, some directionality has been noted; the extensive observations reported by \citet{Ponti10} strongly suggest an overall movement toward positive Galactic longitudes, away from the vicinity of Sgr A*.  That, vicinity, however, contains far more than just the central black hole.  Within 10 projected parsecs of Sgr A*, there are pulsar wind nebulae, supernova remnants, and presumably vast numbers of stellar remnants.  Indeed, \citet{Fryer06} have suggested that a powerful X-ray pulse having a 2 - 200 keV luminosity of $\sim$10$^{39}$ ergs could have been produced by the impact of the Sgr A East supernova remnant upon the adjacent ``50 km s$^{-1}$ molecular cloud.''  They estimate the age of the remnant to be 1700 years, and that the impact began about 500 years ago and has now subsided to a much lower luminosity.  The weakness of this particular hypothesis for the origin of the X-ray flash is the 8-year time scale for the decay of the scattered hard X-ray emission from Sgr B2 reported by \citet{Terrier10}.  Because of the finite size of the Sgr B2 cloud, this is the maximum time scale for the decay of the actual illuminating flare.   A supernova remnant 3 pc in radius impacting an extended cloud seems unlikely to be able to account for such short time-scale variability.  Another model for the production of the X-ray flash -- one that might satisfy the variability time scale constraint -- is that the X-rays were produced in a shock arising from the impact of a jet from Sgr A* upon the dense interstellar medium of the Galactic center \citep{Yu11}.  Yu et al. suggest that the jet arose as a consequence of the tidal disruption of a star.  

If Sgr A* is the direct source of the flash, then an enhanced accretion event is the likely cause.  \citet{Koyama96} raised the possibility that the flash resulted from the tidal capture of a star, and that would easily supply the necessary energy, but given the mean expected time between such captures, 10$^4$ - 10$^5$ years, it requires that we be viewing the Galactic center arena in a rather fortuitous time interval, with an {\it a priori} probability of a few percent or less.    Accretion of smaller bodies of planetary mass could also provide the necessary energy, but the frequency of such events cannot be reliably estimated.  Accretion of interstellar material is perhaps a more frequent occurrence because of their larger cross-section; the vicinity of Sgr A* contains several ionized gas streams \citep[{\it e.g.},][]{Zhao09} and a number of blobs of dust and gas \citep[see also Section \ref{sect:blob}]{Ghez:2005uq,Muzic:2010fk}.  Further study of their trajectories and masses is necessary to assess the likelihood of a sufficiently energetic accretion event.  

\begin{figure}
    \centering
        \includegraphics[width=8cm]{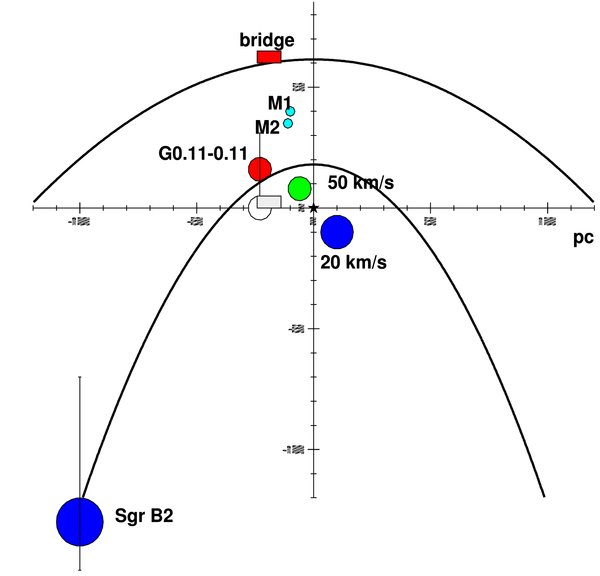}

        \vspace{-3mm}
        \caption{Face-on view of the Galactic disk as seen from the north Galactic pole. Sgr A* is the black star at the origin. The direction of the Earth is toward the bottom of the figure.  Possible line-of-sight placements of various clouds are indicated in color, and the white versions are at the projected distances. All are currently fluorescing in the 6.4 keV line except the 20 and 50 km/s clouds.  Sgr B2 is shown as a blue circle about 130 pc in front of Sgr A*, the distance estimated by \citet{Reid09SgrB2} using trigonometric parallax. The parabolas shown represent the surfaces that are simultaneously illuminated, {\em from our terrestrial perspective}, by light fronts that were emitted 100 (bottom) and 400 (top) years ago from Sgr A*. From \citet{Ponti10}.}
       
        \label{echo}
   \end{figure}

\subsection{duration and multiplicity of flashes}

If the temporal profile of the hard X-ray flash were precisely known, the timing of the appearance and disappearance of the fluorescent signal from individual clouds could be used to locate those clouds along the line of sight.  Such tomography of the central molecular zone would provide a major advance in our understanding of gas dynamics in the Galaxy's central bar potential.  
However, this ideal may take some time to realize.  Figure~\ref{echo}, 
from \citet{Ponti10}, illustrates how several quite different possibilities remain.   The flash could have had a duration as long as several hundred years, it could have been a single brief flash, or it could have consisted of a train of several flashes emitted over several hundred years.  All that we presently know is that at least one of the flashes faded quickly, and that there is no evidence that any flashes have occurred in the last 100 years or so. Future progress on simultaneously constraining the temporal profile of the flash and the placement of clouds will require a decades-long monitoring effort.  
   
However, a novel way of elucidating the scattering geometry has recently been suggested by \citet{Capelli12}.  Using a global average of XMM data acquired between 2000 and 2009 of a region within $\sim$30-pc of Sgr A*, these authors use the intensity ratio of the scattered hard X-ray continuum to the 6.4 keV neutral iron line to infer an iron-based metallicity of 1.7 times the solar value.  They then note that the scattering of the X-ray continuum is angle-dependent, while the K$\alpha$ line emission following the ejection of the iron K electron is isotropic.  Consequently, if the iron abundance is known, or assumed to be fixed at their determined value, the X-ray line-to-continuum ratio can be used as a measure of the scattering angle, and thus the line-of-sight distance, modulo a sign ambiguity.  \citet{Capelli12} thereby offer preliminary estimates of the line-of-sight displacements of some clouds from Sgr A*.  It is not clear, however, that they have accounted for the differential penetration into the clouds of the K-shell ionizing photons and the adjacent X-ray continuum, which \citet{Odaka11} argue can also give rise to an angle-dependent line-to-continuum ratio.  
 
 \subsection{X-ray production of gas-phase SiO}
 
\begin{figure}
    \centering
        \includegraphics[width=9cm]{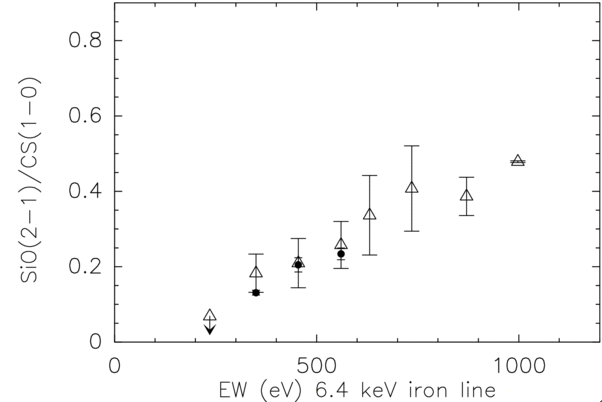}

        \vspace{-3mm}
        \caption{Correlation between the SiO(2-1)/CS(1-0) line intensity ratio and the equivalent width of the 6.4 keV neutral iron line.   The open triangles represent averages of many positions in bins of 100 eV, except for the two points with the highest EW, which were averaged together. Error bars show the 1 $\sigma$ standard deviation of the averaged data. Filled circles in three of the bins result from the inclusion of upper limits.  From \citet{AmoBaladron09}.}
       
        \label{Arancha}
   \end{figure}

 The chemistry of the clouds in the central molecular zone also appears to be affected by the passage of the X-ray flash.  \citet{MartinPintado00} pointed out a correlation between the distribution of rotational line emission from SiO and the distribution of 6.4 keV line emission.  Since SiO is generally a refractory molecule and silicon tends to reside predominantly in grains in the form of silicate compounds, this was an indication that either the X-rays, or something associated with them, are removing the silicon from grains in the form of SiO.  This work was followed up by \citet{AmoBaladron09}, who compared 6.4 keV maps produced from Chandra data with the intensity ratio of SiO J = 2-1 to CS J = 1-0 line emission in a broad region near Sgr A*.  CS was used for comparison, because it is a predominantly gas-phase molecule that is collisionally excited in the same range of densities as SiO.  These authors confirmed and strengthened the earlier correlation (figure \ref{Arancha}), and suggested that the release of SiO from small grains can occur when hard X-rays deposit their energy in the grains, and heat them to temperatures as high as 1000 K.  In that sense, the regions displaying excess SiO are X-ray dominated regions, or XDRs.  Further work is now need to investigate other potential chemical diagnostics of XDR conditions.  
 
The X-ray flashes might also play a significant role in cloud heating.  The temperatures of most Galactic center clouds are quite high ($\sim$80 K) compared to clouds in the Galactic disk.  \citet{RodFer04} examined potential mechanisms for cloud heating in the Galactic center, and concluded that X-ray heating is not competitive with other mechanisms.  However, they examined present-day X-ray luminosities for this assessment, whereas an X-ray flash several orders of magnitude more intense than any current source in the Galactic center could have contributed substantially to cloud heating.  If the time since such a flash were shorter than the cloud cooling time, then the influence of the X-ray flashes could be persistent, and possibly even dominant.  
   
\subsection{Concluding remarks on the neutral iron K$\alpha$ emission}  
In conclusion, the case for the X-ray fluorescence mechanism for the bulk of the 6.4 keV line emission now appears incontrovertible.  However, there is still room for energetic particle production of the iron line in particular environments.  The Arches Star Cluster, for example, could produce a local cosmic ray flux that accounts for the fluorescent X-ray emission in its vicinity \citep{Capelli11}.  In addition, cosmic ray protons permeating the central molecular zone could be producing a persistent or very slowly varying neutral Fe K$\alpha$ signal.  In examining this hypothesis, \cite{Dogiel11} suggest that when the emission caused by the hard X-ray continuum fades out as the flash passes through a cloud, there could remain a weaker, quasi-stationary component generated by proton impact.   The two distinct processes can be distinguished using equivalent width as a diagnostic, because they differ in their relative production of X-ray continuum and iron-line fluxes.  


\section{VLBI (sub-)millimeter imaging of Sgr~A* -- on the path to resolving the shadow}

In General Relativity, photons move along geodesics in a curved spacetime, leading to the bending of light rays. Therefore,  the appearance of a black hole put in front of a larger bright source is expected to be a black disk with a diameter greater than the event horizon. This effect has been called the "shadow" of the black hole \citep{falcke00}. In 1974, Bardeen described this phenomenon and noted: "Unfortunately, there seems to be no hope of observing this effect." \citep{bardeen74} Almost 40 years later, there is good reason for hope. The estimated angular diameter of the black disk for Sgr~A* is in a regime that future  VLBI measurements conducted  at the shortest wavelengths possible should be able to resolve. Fortunately, at these sub-mm wavelengths the interstellar scattering starts to become negligible and Sgr~A*'s accretion flow is likely optically thin, both of which are crucial ingredients for the quest to image  the shadow of a black hole. 

\subsection{The concept of a black hole shadow}

Although the black hole at the Galactic center is surrounded by an accretion flow that emits radiation, it is nevertheless  instructive to look first at a case where the flux is only coming from a source behind the black hole. The question that needs to be investigated is which photons will ultimately end up within the horizon and which make it to the observer at infinity. The geodesics of the Schwarzschild metric, which describes a non-spinning black hole, and of the Kerr metric, which describes a rotating black hole, are well understood \citep[e.g.][]{bardeen}. In the non-rotating case, the cross section for gravitational capture of a photon is given by a circle with radius $r = \sqrt{27}/2 \cdot R_S$, with $R_S = 2GM/c^2$ being the Schwarzschild radius. This evaluates to $\sim$25 $\mu$arcsec on sky for Sgr~A*.  In the rotating case, when the black hole spins with maximum angular momentum, the shape deviates from a circle and the radius of the cross section for the photons that propagate perpendicular  and parallel to the black hole's spin axis are $r_{\perp} = \sqrt{24.3}/2 \cdot R_S$ and $r_{\parallel} = \sqrt{23.3}/2 \cdot R_S$, respectively \citep{young76}. Hence, the shadow  should not vary dramatically with the black hole's spin.

The above considerations should also hold approximately true for the case that the black hole is surrounded by an optically thin emission source. The reason for this is that the flux coming from the near side of the black hole is suppressed relative to the flux coming from the far side for two reasons: (i) strong gravitational redshift and a shorter total path length (meaning smaller integrated emissivity) decrease the emission from the near side, and (ii) gravitational lensing amplifies the radiation from the far side \citep{falcke00}.  

\begin{figure}
    \centering
        \includegraphics[width=13cm]{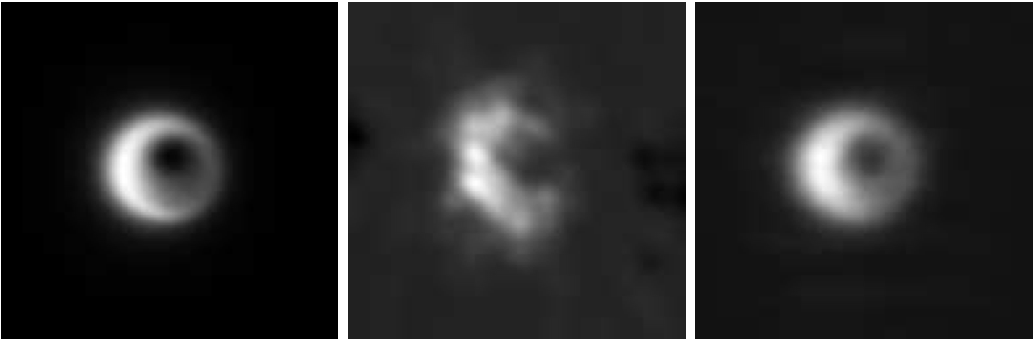}

        \vspace{-3mm}
        \caption{Simulations of the potential appearance of Sgr~A* at a wavelength of 0.8\,mm. The left panel shows the image as it would be seen by a perfect telescope; the blurring is due to interstellar scattering. The middle panel shows an image as it could be reconstructed with a planned seven VLBI stations array (see Figure~\ref{stations}). The right panel shows the reconstructed image obtained with a simulated array consisting of 13 stations. The central depletion of flux is called the shadow of the black hole. It can potentially be observed with seven future sub-mm VLBI stations. The inclination in the simulations was set to 30\degr and the black hole is assumed to be non-rotating. From \cite{fish09}.}
        \label{shadow}
   \end{figure}

A more realistic answer for how an image of Sgr~A* in the sub-mm regime might appear, however, needs to be found using simulations that employ detailed accretion dynamics and viewing angle effects. Figure~\ref{shadow} shows an example of such a simulation. A depletion of flux near the center is clearly visible. The asymmetry in the flux distribution is caused by the Keplerian rotation of the accretion flow, which leads to relativistic beaming. The appearance of a shadow seems to be a robust feature and this also holds true in situations where the accretion flow's angular momentum and the black hole's angular momentum axes are not aligned \citep{dexter12}. 

The image of a black hole's silhouette would mark a crucial visualization and test of strong-field gravity effects. However, it is not the event horizon that is causing the shadow and the sub-mm VLBI measurements alone cannot resolve the question of whether the Galactic center black hole truly does have an event horizon, the defining property of a black hole.  \cite{broderick06} and \cite{broderick09} show that when these data are combined with near-infrared observations of Sgr~A*, severe constraints can be placed on the size and luminosity of a putative surface. Since the surface's emission would have to approach that of a blackbody with maximum intensity in the near-infrared, the observations limit the accretion flow's radiative efficiencies. Although current limits are already deemed to be in violation of the observed accretion efficiencies, future sub-mm VLBI should further strengthen the argument, thereby possibly all but requiring an event horizon. 

\subsection{Current mm-VLBI observations of Sgr~A*}

\begin{figure}
    \centering
        \includegraphics[width=8cm]{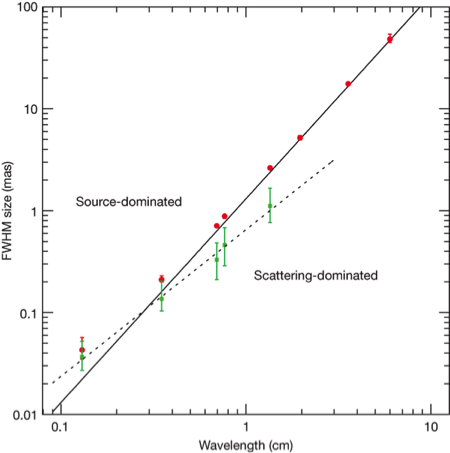}

        \vspace{-3mm}
        \caption{The observed size of Sgr~A* as a function of wavelength. Red points show major-axis observed sizes, and green points show derived major-axis intrinsic sizes of Sgr A*. The solid black line is the best-fit interstellar scattering law as derived from measurements made at wavelengths $> 17$\,cm. Below this line, measurements of the intrinsic size of Sgr~A* are dominated by scattering effects, while measurements above the line indicate structures that are intrinsic to Sgr~A*. This shows that observations below $\sim$2\,mm can resolve the source itself. From \cite{doeleman08}.}
        \label{scatter}
   \end{figure}

VLBI observations of Sgr~A* offer the highest angular resolution possible today, but they suffer from interstellar scattering which leads to a broadening of the radio image. For theoretical reasons a $\lambda^2$ dependence of the interstellar scattering is expected, and indeed such a dominant variation of the angular size was observed up to 43 GHz ($\lambda = 7$\,mm, see Fig.~\ref{scatter}). Observations at shorter wavelengths, though, promised the unambiguous detection of the intrinsic size of the radio source, Sgr~A*, as the scattering size scales with $\lambda^2$, the intrinsic source scales with $< \lambda^2$, and the telescope resolution behaves like $\lambda$ \citep{bower04}. But poor telescope performance and decreased interferometric coherence make sub-mm VLBI an extremely challenging task. An additional disadvantage for observations of the Galactic center, which is at a declination of $-29\degr$, is that most radio telescopes are located on the northern hemisphere.

Nevertheless, with the data at hand today an intrinsic source size can be measured. This is done via quadrature subtraction, i.e. $\theta_{\mbox{\tiny int}}=\sqrt{\theta^2_{\mbox{\tiny obs}}-\theta^2_{\mbox{\tiny scat}}}$. Consequently, the exact form of the assumed scattering law ($\theta_{\mbox{\tiny scat}}=a \lambda^\beta$) determines $\theta_{\mbox{\tiny int}}$ so that the derived sizes are dependent on the form and normalization of the scattering extrapolation. With the assumption that the scattering law is determined accurately at wavelengths longer than $\sim17$\,cm, the intrinsic size at 3.5\,mm of Sgr~A* could be determined to be $13\, R_S$ \citep{bower06}. Recently, \cite{doeleman08} resolved Sgr~A* at 1.3\,mm, which marked the first measurement in a wavelength region where the data are dominated by the intrinsic source size (see Fig.~\ref{scatter}) . 

While \cite{doeleman08} detected correlated flux from Sgr~A*, they have too few visibility measurements to form an image. Instead, models are fit to the visibilities. Assuming a circular Gaussian source structure and removing the scattering effects, they arrive at an intrinsic source size of $37^{+16}_{-10}$ $\mu$arcsec ($3\sigma$ errors), which corresponds to $3.7\, R_S$. This is quite remarkable, since we have noted above that the expected size of the black hole shadow has a diameter of $5.2\, R_S$. Hence, the observed source cannot be centered on the dark mass, but must be shifted away from the black hole center. It seems likely that the effects of Doppler shift and beaming are responsible for this shift. 

A more physically motivated model of a radiatively inefficient accretion flow has been fitted to the \cite{doeleman08} data by \cite{broderick09b}. The spin and orientation angles of the black hole are kept as free parameters. However, the data are not yet powerful enough to significantly constrain these; only an inclination angle of less than $30\degr$  ($0\degr$ means face-on) can be ruled out. The biggest challenge in this approach is the systematic uncertainty of the accretion flow structure around the Galactic center black hole.

\subsection{The Event Horizon Telescope}

The mm-VLBI stations used for the observations by \cite{doeleman08} were the Arizona Radio Observatory 10-m Submillimeter Telescope (ARO/SMT) on Mount Graham in Arizona, one 10-m element of the Combined Array for Research in Millimeter-wave Astronomy (CARMA) in Eastern California, and the 15-m James Clerk Maxwell Telescope (JCMT) near the summit of Mauna Kea in Hawaii. They reported that Sgr A* was robustly detected on the short ARO/SMT--CARMA baseline and the long ARO/SMT--JCMT baseline. In order to achieve a less model dependent result and -- as the final goal -- an image, more baselines, better {\it uv}-plane\footnote{The {\it uv}-plane is the Fourier transform of the image plane and each of the 2-dimensional baselines corresponds to a sample point within that plane.} coverage, and therefore more sub-mm VLBI stations are needed.

The entirety of existing and planned VLBI stations employed in Sgr~A* imaging is called the "Event Horizon Telescope" (which is somewhat misleading, since it is not necessarily the event horizon that will be observed).  In the near future, the participants are planning to extend this array by including the Atacama Submillimeter Telescope Experiment (ASTE) or Atacama Pathfinder Experiment (APEX) in Chile, the Large Millimeter Telescope (LMT) in Mexico, the Institut de Radioastronomie Millimetrique (IRAM) 30 m telescope at Pico Veleta in Spain, and the IRAM Plateau de Bure interferometer in France. What a Sgr~A* image at a wavelength of 0.8\,mm from this future seven-station VLBI network could look like is shown in the middle panel of Figure~\ref{shadow}. Since also using these seven telescopes leaves large parts of the {\it uv}-plane unsampled, the natural goal is to expand the set of available sub-mm telescopes to at least 13.

\begin{figure}
    \centering
        \includegraphics[width=8cm]{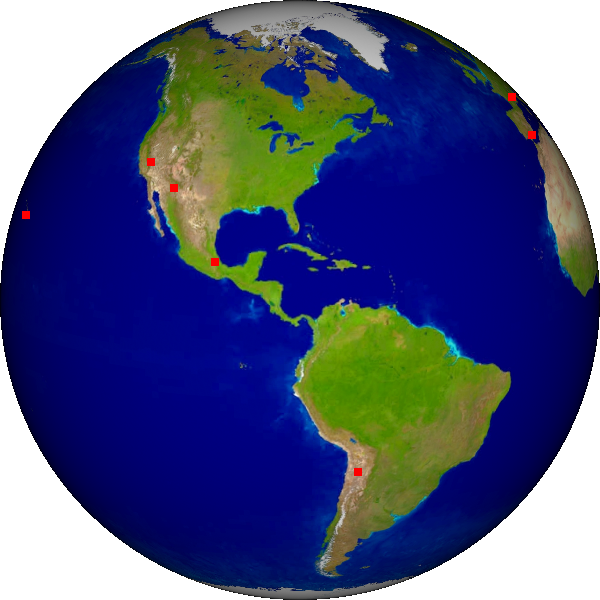}

        \vspace{-3mm}
        \caption{The seven stations that are planned to conduct sub-mm VLBI. Sgr~A* has already been detected at 1.3\,mm with the Hawaii, Arizona, and California stations. The Chile station is crucial to provide the necessary north-south baseline. From www.eventhorizontelescope.org.}
        \label{stations}
   \end{figure}


\section{Measuring the central gravitational potential through stellar orbits}

\begin{figure}
    \centering
        \includegraphics[width=10cm]{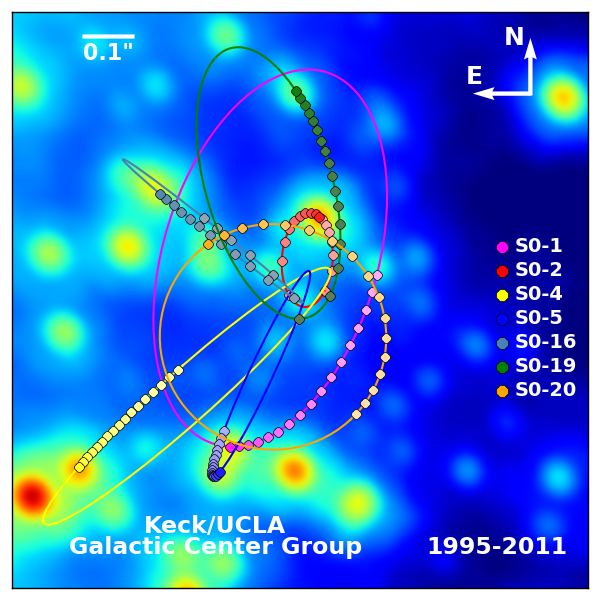}

        \vspace{-3mm}
        \caption{Stellar orbits at the Galactic center proving the existence of a black hole. The star S0-2 dominates our knowledge about the central potential, since with an orbital period of 16 years it has been tracked throughout a whole orbit. The other stars have longer orbital periods and therefore only a fraction of their orbits is covered by observations.}
        \label{orbits}
\end{figure}

High-angular resolution imaging of our Galactic center has identified a dense star cluster at the heart of the Galaxy. The innermost stars -- the so-called S stars -- form a cluster of predominantly young (B-type) stars
orbiting isotropically within $\sim$0.04 pc of the central black hole, and collectively having relatively large eccentricities \citep{Ghez05a,gillessen09a}.  The current paradigm for their dynamical state is that they result from the capture of individual stars by the tidal disruption of binaries that have been scattered close to the central black hole \citep{Hills88,GouldQuillen03,Perets07,Perets09,Fabio10}.  Because of their proximity to the central black hole, they are ideal test-bodies of the gravitational potential in that region. In fact, these stars move at such high velocities (up to $\sim$10,000 km s$^{-1}$) that individual orbits can be derived (see figure~\ref{orbits}). 

The best probe of the Galaxy's central potential to date is the star
S0-2.  It is the star with the shortest known orbital period ($P = 16$ yrs) whose 
orbit can be measured both astrometrically and spectroscopically (it has a magnitude of $K = 14$).  
Measurements of this star have offered us the best evidence to date for the existence of 
a supermassive black hole at the center of our Galaxy \citep[e.g.,][]{schoedel02,ghez03,ghez08,gillessen09a,gillessen09b}.  An analysis of our current data for S0-2's orbit puts the best estimate of the black hole's mass at 4.1 $\pm$ 0.4 $\times$ 10$^6~M_{\odot}$.

Ever since the discovery of S0-2 and other similarly fast-moving stars
at the center of the Milky Way \citep{eckart97,eckart02,ghez98,ghez00}, the prospect of using stellar orbits to 
make ultra-precise 
measurements of the distance to the Galactic center (R$_o$), 
and, more ambitiously, to measure
post-Newtonian effects has been anticipated 
by many under the assumption that radial velocities and more accurate astrometry
would eventually be obtained \citep[e.g.][]{jaro98,rubilar01,weinberg05,zucker06,will08,angelil10,merritt10}.  

\begin{figure}
    \centering
        \includegraphics[width=13cm]{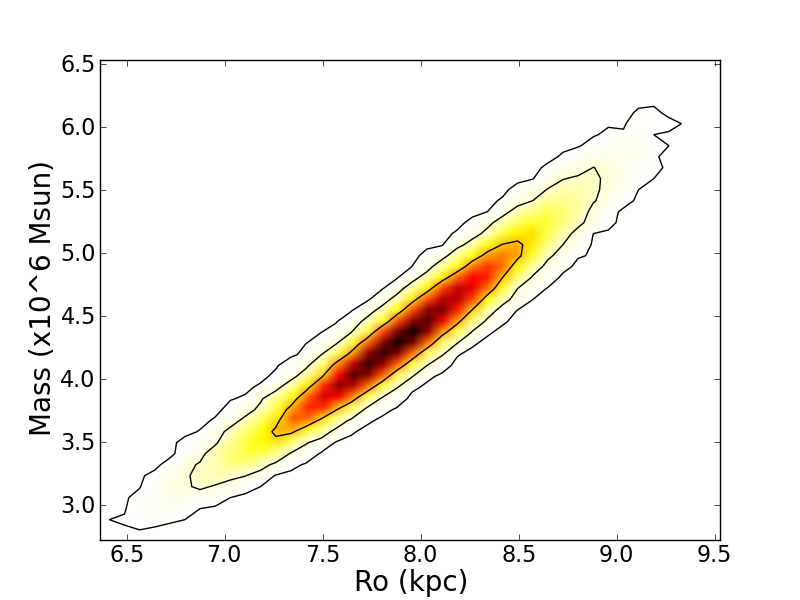}

        \vspace{-3mm}
        \caption{The correlation of the black holeÕs estimated mass and distance. The contours mark the 68\%, 95\%, and 99.7\% confidence limits. While mass and distance are well determined from the orbit of S0-2, they are not independent quantities.}
        \label{massRo}
\end{figure}

R$_o$ -- a fundamental 
parameter for models of the Galaxy --
affects almost all questions of Galactic structure and mass.  
It can be estimated by fitting S0-2's orbital
motions with a Keplerian model \citep[e.g.,][]{ghez08,gillessen09a,gillessen09b}, and our current best fit value for this parameter is 7.7 $\pm$ 0.4 kpc, with an uncertainty that is highly 
correlated with the uncertainty in the mass of the black hole (see figure~\ref{massRo}). 
With continued astrometric and spectroscopic monitoring, the precision in R$_o$ 
should improve by an order of magnitude.
With this measurement accuracy, it would be possible
to improve estimates of the amount and distribution of dark mass in
the Milky Way 
and to possibly recalibrate rungs of the cosmic distance ladder \citep[e.g.][]{olling00,majewski06}.
 
With care, the planned astrometric and spectroscopic 
adaptive optics observations for R$_o$, along with measurements of other short-period stars, 
should enable measurements of 
stellar orbits with sufficient precision to detect 
non-Keplerian effects due 
to General Relativity (GR) and extended dark mass.
The leading order effects are the special relativistic transverse Doppler shift and the 
gravitational redshift. These effects should be observable when S0-2 goes through its next closest approach in 2018 \citep{angelil11}, thereby marking a test of Einstein's equivalence principle. 
 
\begin{figure}
    \centering
        \includegraphics[width=10cm]{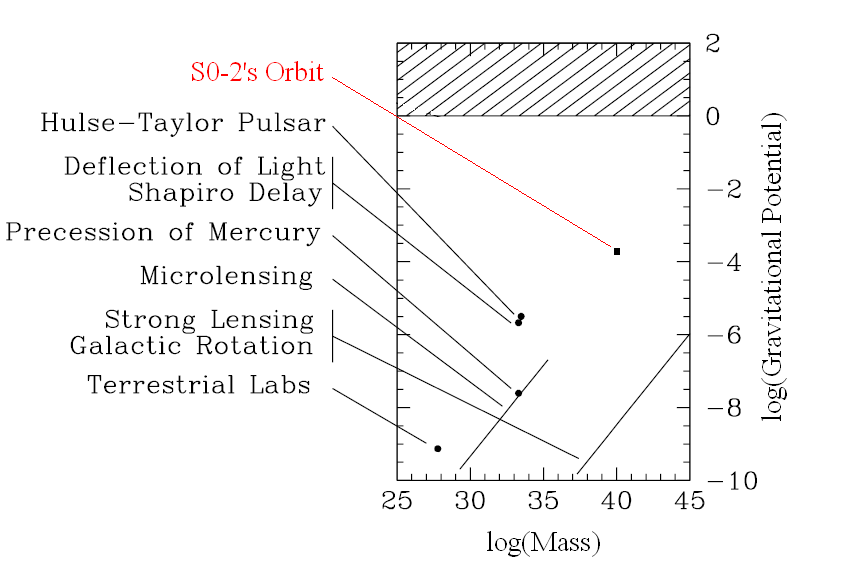}

        \vspace{-3mm}
        \caption{The most important tests of GR in context with the orbit of S0-2, which could be used to probe gravity. The gravitational potential -- a measure of how strong a field is (at $\sim$1, the event horizon is reached) -- and the mass of the source are used as parameters.  S0-2Õs orbit provides access to a new regime. Figure adopted from \citet{psaltis04}.}
        \label{tests}
\end{figure}

An unambiguous detection of the prograde general relativistic precession of the periapse is more involved, since it is degenerate with retrograde Newtonian effects caused by extended mass.  Current assumptions about the distribution of dark mass within several thousand AU of the Galactic black hole yield a prediction for the relativistic precession of the periapse for S0-2 that is, in total, about twice as large in magnitude as that due to the extended mass terms.
A detection of GR effects in the orbital motions of the
short-period stars at the Galactic center would probe
an unexplored regime of GR, which is the least tested of the 
four fundamental forces of nature (figure~\ref{tests}).  
Ground-based interferometric measurements with future instrumentation
of gas motions much closer in to the black hole may also offer
another probe of GR (see Genzel et al. 2010), although this measurement
depends on the notion that observationally isolatable gas 
concentrations exist and are stable for extended 
periods of time close to the innermost stable circular orbit.

\section{Fast-moving red emission-line source -- dust cloud or stellar in nature?}
\label{sect:blob}

\begin{figure}
    \centering
        \includegraphics[width=13cm]{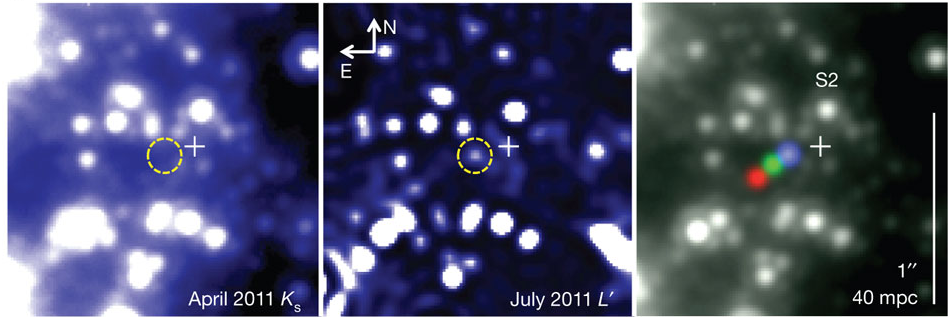}

        \vspace{-3mm}
        \caption{VLT/NaCo observations of the center of the Galaxy. The circle marks the position of the putative gas cloud, and the cross marks the position of Sgr~A*. The left panel shows a K-band (2.2$\mu$m) image. At this wavelength, the emission-line source is not detected. At L' (3.8$\mu$m) on the other hand, the source is detected (middle panel). The right panel shows its motion towards Sgr~A* (the colored points are its positions in 2004.5 (red), 2008.3 (green), and 2011.3 (blue). From \cite{gillessen12}.}
        \label{cloud}
   \end{figure}

Recently, Gillessen et al.~(2012) reported the discovery of a 3 $M_{ {Earth}}$ cloud of gas and dust on its way towards the supermassive black hole in the Galactic center. The
identification of the source as a dusty cloud is based on 
its Br-$\gamma$ emission and its very red colors (see Figure~\ref{cloud}).  The source is rapidly accelerating and is predicted to
approach the black hole at a distance of only 3100 times the radius of the event horizon in
mid-2013. The gas cloud would likely be torn apart by the tidal forces
of the black hole and then accreted. In this scenario, the accretion flow onto the Galactic black hole is expected
to be temporarily disturbed by the added mass of this cloud, thereby giving rise to an enhanced accretion luminosity. The encounter of the gas cloud with the black hole could be an ideal
experiment to probe the physics of accretion onto a supermassive black hole.

The attribute cited by Gillessen et al.~(2012) as the foundation of their gas blob interpretation is that the Br-$\gamma$ emission is spatially resolved, growing larger, and increasing its velocity dispersion between 2008 and 2011. This would be consistent with theoretical simulations of the tidal disruption of a small unbound gas cloud, in which the cloud stretches along its orbital path as it approaches the black hole. 


However, there are a number of problems with the gas cloud hypothesis.
The main challenge is to explain how such
a small, low-mass gas cloud is able to stay intact in the
gravitational potential of the black hole. Using the assumption of a gas
cloud, estimates of its temperature and Br-$\gamma$ flux places its
mass at about 3 $M_{\it Earth}$. As this mass is too low to be bound by
self-gravity given its size, the gas cloud
is thought to be confined by ram pressure with the surrounding gas.
However, Rayleigh-Taylor instabilities at the gas cloud interface should break up
the cloud within a few years (Morris~2012), leading Gillessen et al.~(2012) to propose that the gas
cloud originated very locally, perhaps in the colliding winds of nearby
massive stars.

\begin{figure}
    \centering
        \includegraphics[width=13cm]{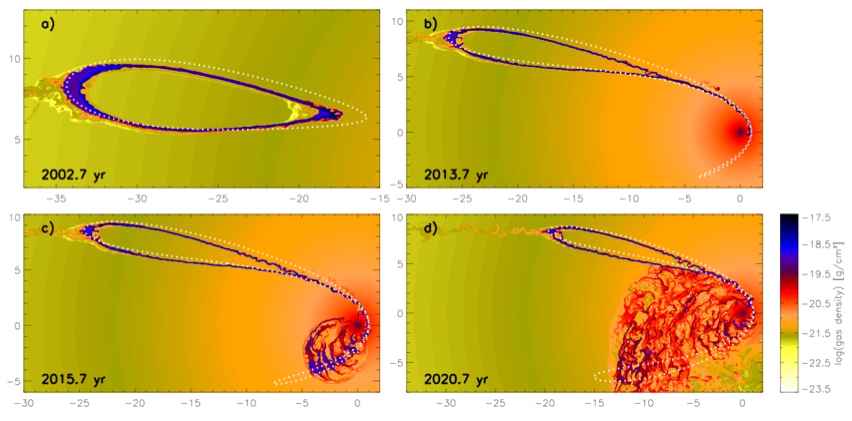}

        \vspace{-3mm}
        \caption{The preferred model, assuming the source is a low-mass gas blob, is not a compact cloud, but rather an extended shell structure. Since a compact gas cloud could not survive long in the Galactic Center environment, and its hydrodynamic modeling does not reproduce all observational characteristics, \cite{burkert12} prefer a spherical shell scenario in which the observed source is the head of a much larger ring-like structure, the density evolution of which is shown. From \cite{schartmann12}}
        \label{shell}
   \end{figure}

\cite{burkert12} and \cite{schartmann12} present hydrodynamical simulations of a compact gas cloud in the environment of the Galactic center. They confirm that such a cloud would have had to be created very recently, in 1995 according to their analysis. This is only a few years before the adaptive optics instrument NaCo enabled the cloud's first detection in 2002. In this interpretation, it seems to be a remarkable coincidence that the gas cloud popped into existence at around the same time when its observation became possible.  This is one of the reasons why the authors prefer a second hypothesis: the spherical shell scenario, in which the source is interpreted as the head of an already disrupted spherical shell or ring (see Figure~\ref{shell}). Such a structure could be older, but it is not obvious why a shell like this should be appearing. 

While the source does show several unusual properties compared with most stars in the Galactic center -- it is highly reddened, with detections in the L'-band (3.8$\mu$m) and the M-band (4.7$\mu$m), but it is not detected in the K-band (2.2$\mu$m; Gillessen et al.~2012), and it shows strong Br-$\gamma$ emission -- none of these attributes can distinguish between a stellar source, such as a Be-star, with a disk or dusty envelope and a gas blob. We want to note that we detect several sources with similar characteristics throughout the Galactic center region. 

An alternative hypothesis for the nature of this very red emission-line source is that it is circumstellar. The observed Br-$\gamma$
emission-line and limits on its infrared colors are consistent with
that observed from Herbig Ae/Be and classical Be stars. Herbig Ae/Be
stars are found in very young star forming regions, while classical Be
stars begin to appear at about 5 Myr \citep{grebel97}. Given the age of
the young star cluster in this region (4-6 Myr), we may expect to
observe a few such stars among the $>70$ B stars found so far.

\begin{figure}
    \centering
        \includegraphics[width=8cm]{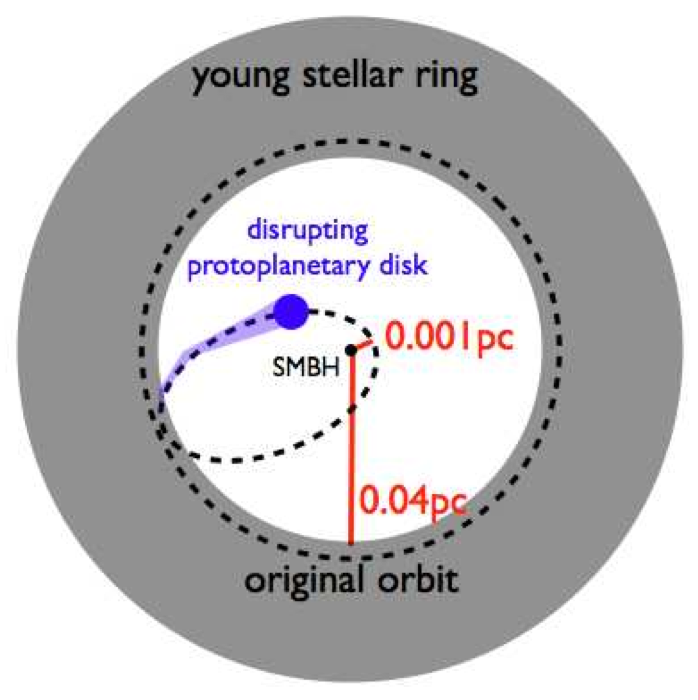}

        \vspace{-3mm}
        \caption{A sketch of the scenario proposed by \cite{loeb12}. A young, low-mass star that is surrounded by a proto-planetary disk gets scattered away from the inner edge of the disk of young stars and towards the black hole. The circumstellar disk gets tidally disrupted and photoevaporated, reproducing the observed size of the gas cloud. The implications of this stellar model for future observations are quite distinct in contrast to the model of a gas shell, which is expected to get completely disrupted in the coming years.}
        \label{disk}
   \end{figure}

A proto-planetary stellar disk origin of the red emission-line source has been proposed by \cite{loeb12}. They suggest that the gas cloud surrounds a low-mass star, which was scattered away from its original low-eccentricity orbit at the inner edge of the disk of young O/WR stars that is known to orbit the galactic black hole \citep{bartko09, lu09, yelda12}. While the star itself is hypothesized to be of too low mass to be observable, the debris produced through the disruption of its proto-planetary disk allows it to be detected. The disruption and therefore the observed size of the gas cloud is caused in this model by photoevaporation and tidal stretching of the disk (Figure~\ref{disk}). 
The authors argue that while it seems {\it a priori} unlikely that a star gets deflected from an orbit with small eccentricity in the disk of young stars to the observed highly eccentric orbit without disrupting the gaseous disk, it is nevertheless plausible given the observations.

\cite{escude12} also proposes a circumstellar gas disk as an interpretation of the data. However, in this scenario the star is old. The author suggests that a disk was formed when an old, low-mass star suffered a close encounter with a stellar mass black hole. This can tidally disrupt  the star's outer envelope and deflect the star onto the observed orbit. Even though some of the tidal debris may have escaped the star, a large fraction of the mass stayed bound and fell back to the star, creating a small disk. 


Luckily, observations in the short-term future will likely settle the question of the nature of the red emission-line source as it swings around the Galactic black hole in 2013, since the evolution of a gas cloud around a low-mass star should be easily distinguishable from that of a pressure-confined gas cloud with no self-gravity or central mass supply.   In any case, this object has provoked considerable interest in what the observable manifestations of an episodic increase in the accretion rate onto the black hole might be.  For example, \citet{MSGD12} have pointed out that modest increases in the accretion rate would have important implications for the Event Horizon Telescope: the characteristic ring of emission due to the photon orbit becomes brighter, more extended, and easier to detect, and the emission at all observable wavelengths can brighten considerably.  In addition, many observations are now being proposed and planned to monitor Sgr A* in anticipation that the show could turn out to be quite interesting.

\normalem

\begin{acknowledgements}
This work was funded by the US National Science Foundation under grant AST 09-09218 to UCLA.  
\end{acknowledgements}


\label{lastpage}


\begin{thebibliography}{99}
\small \setlength{\itemindent}{-3mm} \setlength{\itemsep}{-0.5mm}
\setlength{\baselineskip}{4.8mm}

\bibitem[Amo-Baladr\'on {\it et al.}(2009)]{AmoBaladron09} Amo-Baladr\'on, M.A., Mart\'in-Pintado, J., Morris, M.R., Muno, M.P., Rodr\'iguez-Fern\'andez, N.J. 2009, ApJ 694, 943.

\bibitem[Antonini {\it et al.}(2010)]{Fabio10} Antonini, F., Faber, J., Gualandris, A., Merritt, D. 2010, ApJ 713, 90.

\bibitem[Ang\'elil {\it et al.}(2010)]{angelil10}  Ang\'elil, R., Saha, P., Merritt, D. 2010, \apj, 720, 1303.

\bibitem[Ang\'elil \& Saha(2011)]{angelil11} Ang\'elil, R.,  Saha, P. 2011, \apj , 734, L19.

\bibitem[Baganoff {\it et al.} (2001)]{Baganoff01} Baganoff, F.K., Bautz, M.W., Brandt, W.N. {\it et al.} 2001, Nature 413, 45.

\bibitem[Baganoff {\it et al.}(2003)]{Baganoff03} Baganoff, F.K., Maeda, Y., Morris, M., {\it et al.} 2003, ApJ 591, 891.

\bibitem[Bardeen {\it et al.}(1972)]{bardeen} Bardeen, J.~M., Press, W.~H., Teukolsky, S.~A. 1972, ApJ, 178, 347.

\bibitem[{Bardeen}(1974)]{bardeen74}
Bardeen, J. M. 1974, In: Gravitational radiation and gravitational collapse; Proceedings of the Symposium, Dordrecht, D. Reidel Publishing Co.,  p. 132-144

\bibitem[Bartko {\it et al.}(2009)]{bartko09}
Bartko, H. et al. 2009, \apj, 697, 1741. 

\bibitem[Becklin \& Neugebauer(1968)]{BN68} Becklin, E.E., Neugebauer, G. 1968, ApJ 151, 145.

\bibitem[Bower {\it et al.}(2002)]{bower02} Bower, G.C., Falcke, H., Sault, R.J., Backer, D.C. 2002, ApJ 571, 843. 

\bibitem[Bower {\it et al.}(2004)]{bower04} Bower, G. C., Falcke, H., Herrnstein, R. M., Zhao, J.-H., Goss, W. M., Backer, D. C., 2004, Science, 304, 704

\bibitem[Bower {\it et al.}(2006)]{bower06} Bower, G. C., Goss, W. M., Falcke, H., Backer, D. C., Lithwick, Y., 2006, ApJ, 648, L127

\bibitem[Broderick \& Narayan(2006)]{broderick06}
Broderick, A. E., Narayan, R. 2006, \apj, 638, L21

\bibitem[Broderick {\it et al.}(2009a)]{broderick09}
Broderick, A. E., Loeb, A., Narayan, R. 2009a, \apj, 701, 1357

\bibitem[Broderick {\it et al.}(2009b)]{broderick09b}	
Broderick, A. E., Fish, V. L., Doeleman, S. S., Loeb, A. 2009b, \apj, 697, 45.

\bibitem[Brown \& Lo(1982)]{BrownLo82} Brown, R.L., Lo, K.Y. 1982, ApJ 253, 108.

\bibitem[Burkert {\it et al.}(2012)]{burkert12} Burkert, A., et al. 2012, \apj, 750, 58.

\bibitem[Capelli {\it et al.}(2011)]{Capelli11} Capelli, R., Warwick, R.S., Porquet, D., Gillessen, S., Predehl, P. 2011, \aap 530, A38.  

\bibitem[Capelli {\it et al.}(2012)]{Capelli12} Capelli, R., Warwick, R.S., Porquet, D., Gillessen, S., Predehl, P. 2012, arXiv:1207.1436.

\bibitem[Chernyshov {\it et al.}(2012)]{Chernyshov12} Chernyshov, D., Dogiel, V., Nobukawa, M., Go Tsuru, T., Koyama, K., Uchiyama, H., Matsumoto, H. 2012, \pasj 64, 14.

\bibitem[Dexter \& Fragile(2012)]{dexter12}
Dexter, J., Fragile, P. C. 2012,  eprint arXiv:1204.4454.

\bibitem[Do {\it et al.}(2009)]{do09} Do, T., Ghez, A. M., Morris, M. R., et al. 2009, ApJ, 691, 1021.

\bibitem[Dodds-Eden {\it et al.}(2009)]{dodds-eden09} Dodds-Eden, K., Porquet, D., Trap, G., et al. 2009, \apj, 698, 676.

\bibitem[Dodds-Eden {\it et al.}(2011)]{dodds11} Dodds-Eden, K., et al. 2011, \apj, 728, 37

\bibitem[Doeleman {\it et al.}(2008)]{doeleman08}
Doeleman, S. S., et al. 2008, Nature, 455, 78

\bibitem[Dogiel {\it et al.} (2009)]{Dogiel09} Dogiel, V. et al. 2009, \pasj 61, 901.

\bibitem[Dogiel {\it et al.}(2011)]{Dogiel11} Dogiel, V. Chernyshov, D., Koyama, K., Nobukawa, M., Cheng, K.-S. 2011, \pasj 63, 535.

\bibitem[Eckart \& Genzel(1997)]{eckart97} Eckart, A., Genzel, R. 1997, MNRAS, 284, 576

\bibitem[Eckart {\it et al.}(2002)]{eckart02} Eckart, A., et al. 2002, MNRAS, 331, 917.

\bibitem[Eckart {\it et al.}(2004)]{eckart04} Eckart, A., Baganoff, F.K., Morris, M., et al. 2004, \aap 427, 1.

\bibitem[Eckart {\it et al.}(2006a)]{eckart06a} Eckart, A., Baganoff, F.K., Sch\"odel, R., et al. 2006a, 450, 535.

\bibitem[Eckart {\it et al.}(2006b)]{eckart06} Eckart, A., Sch\"odel, R., Meyer, L., et al. 2006b, A\&A, 455, 1.

\bibitem[Eckart {\it et al.}(2008)]{eckart08} Eckart, A., Sch\"odel, R., Garc\'ia-Mar\'in, M., et al. 2008, \aap 492, 337.

\bibitem[Eckart {\it et al.}(2009)]{eckart09} Eckart, A., Baganoff, F.K., Morris, M.R., et al. 2009, \aap 500, 935.

\bibitem[Eckart {\it et al.}(2012)]{eckart12} Eckart, A., Garc\'ia-Mar\'in, M., Vogel, S.N., et al. 2012, \aap 537, A52.  

\bibitem[Falcke {\it et al.}(2000)]{falcke00}
Falcke, H., Melia, F., Agol, E. 2000, \apj, 528, L13

\bibitem[Fish \& Doeleman(2009)]{fish09}
Fish, V. L., Doeleman, S. S. 2009, IAU Symposium \#261, eprint arXiv:0906.4040

\bibitem[Fryer {\it et al.} (2006)]{Fryer06} Fryer, C.L., Rockefeller, G., Hungerford, A., Melia, F. 2006, ApJ 638, 786.

\bibitem[Genzel {\it et al.}(2003)]{genzel03}
  Genzel, R., Sch\"odel, R., Ott, T., et al. 2003, Nature, 425, 934.

\bibitem[Genzel, Eisenhauer \& Gillessen(2010)]{GEG10} Genzel, R., Eisenhauer, F., Gillessen, S. 2010, Rev. Mod. Phys. 82, 3121.

\bibitem[Ghez {\it et al.}(1998)]{ghez98} Ghez, A. M., et al. 1998, \apj, 509, 678.

\bibitem[Ghez {\it et al.}(2000)]{ghez00} Ghez, A. M., et al. 2000, Nature, 407, 349.

\bibitem[Ghez {\it et al.}(2003)]{ghez03} Ghez, A. M., et al. 2003, \apj, 586, L127.

\bibitem[Ghez {\it et al.}(2004)]{ghez04} Ghez, A. M., Wright, S. A., Matthews, K., et al., 2004, \apj, 601, L159.

\bibitem[Ghez {\it et al.}(2005a)]{Ghez05a} Ghez, A.M. Salim, S., Hornstein, S.D., Tanner, A., Lu, J.R., Morris, M.R., Becklin, E.E., Duch\^ene, G. 2005a, ApJ 620, 744.

\bibitem[Ghez {\it et al.} (2005b)]{Ghez:2005uq} Ghez, A.M. {\it et al.} 2005b, ApJ 635, 1087.

\bibitem[Ghez {\it et al.}(2008)]{ghez08} Ghez, A. M. et al. 2008, ApJ, 689, 1044.

\bibitem[Gillessen {\it et al.}(2009a)]{gillessen09a} Gillessen, S., et al. 2009a, \apj, 692, 1075.

\bibitem[Gillessen {\it et al.}(2009b)]{gillessen09b} Gillessen, S., et al. 2009b, \apj, 707, L114.
  
\bibitem[Gillessen {\it et al.}(2012)]{gillessen12} Gillessen, S., Genzel, R., Fritz, T. K., et al. 2012, Nature, 481, 51
  
\bibitem[Goldwurm (2011)]{Goldwurm11review} Goldwurm, A. 2011, in Morris, M.R., Wang, Q.D., Yuan, F., eds., ASP Conf. Ser. Vol. 439, The Galactic Center: A Window to the Nuclear Environment of Disk Galaxies, San Francisco, ASP, p. 391.  

\bibitem[Gould \& Quillen (2003)]{GouldQuillen03} Gould, A., Quillen, A.C. 2003, ApJ 592, 935.
  
\bibitem[Grebel(1997)]{grebel97}
Grebel, E. K. 1997, \aap, 317, 448

\bibitem[Herrnstein {\it et al.}(2004)]{Herrnstein04} Herrnstein, R.M., Zhao, J.-H., Bower, G.C., Goss, W.M. 2004, \aj 127, 3399.

\bibitem[Hills (1988)]{Hills88} Hills, J.G. 1988, Nature 331, 687.
  
\bibitem[Inui {\it et al.}(2009)]{Inui09} Inui, T., Koyama, K., Matsumoto, H., Tsuru, T.G. 2009, \pasj 61, 241.
  
\bibitem[Jaroszynski(1998)]{jaro98} Jaroszynski, M. 1998, { Acta Astronomica}, { 48}, 653.
  
\bibitem[Koyama {\it et al.}(1996)]{Koyama96} Koyama, K., Maeda, Y., Sonobe, T., Takeshima, T., Tanaka, Y.,  Yamauchi, S. 1996, \pasj 48, 249.

\bibitem[Koyama {\it et al.}(2008)]{Koyama08} Koyama, K., Inui, T., Matsumoto,H., Tsuru, T.G. 2008, \pasj 60, 201.
  
\bibitem[Lu {\it et al.}(2009)]{lu09}
Lu, J. R. et al. 2009, \apj, 690, 1463.

\bibitem[Macquart \& Bower(2006)]{MacBow06} Macquart, J.P., Bower, G.C. 2006, ApJ 641, 302.

\bibitem[Majewski {\it et al.}(2006)]{majewski06} Majewski, S. R., Law, D. R., Polak, A. A., Patterson, R. J. 2006,
ApJ 637, L25.

\bibitem[Marrone {\it et al.}(2006)]{marrone06} Marrone, D.P., Moran, J.M., Zhao, J.H., Rao, R. 2006, \apj 640, 308.  

\bibitem[Marrone {\it et al.}(2008)]{marrone08} Marrone, D.P., Baganoff, F.K., Morris, M.R., et al. 2008, \apj 682, 373.  

\bibitem[Mart\'in-Pintado {\it et al.}(2000)]{MartinPintado00} Mart\'in-Pintado, J., de Vicente, P., Rodr\'iguez-Fern\'andez, N.J., Fuente, A., Planesas, P. 2000, \aap 356, L5.

\bibitem[Mauerhan {\it et al.}(2005)]{Mauerhan05} Mauerhan, J.C., Morris, M., Walter, F., Baganoff, F.K. 2005, ApJ 623, L25.
  
\bibitem[McHardy {\it et al.}(2006)]{mchardy} McHardy, I. M., Koerding, E., Knigge, C., Uttley, P., Fender, R. P., 2006, Nature, 444, 730.

\bibitem[Melia \& Falcke (2001)]{MeliaFalcke} Melia, F., Falcke, H. 2001, ARAA 39, 309.

\bibitem[Merritt {\it et al.}(2010)]{merritt10} Merritt, D., et al. 2010, \prd,  81, 062002.

\bibitem[Meyer {\it et al.}(2006a)]{meyer06a} Meyer L., Sch\"odel R., Eckart A., Karas, V., Dovciak, M., Duschl, W. J., 2006a, A\&A, 458, L25.

\bibitem[Meyer {\it et al.}(2006b)]{meyer06b} Meyer L., Eckart A., Sch\"odel R., Duschl, W. J., Muzic, K., Dovciak, M., Karas, V., 2006b, A\&A, 460, 15 

\bibitem[Meyer {\it et al.}(2007)]{meyer07} Meyer, L., Sch\"odel, R., Eckart, A., Duschl, W. J., Karas, V., Dovciak, M., 2007, A\&A, 473, 707

\bibitem[Meyer {\it et al.}(2008)]{meyer08} Meyer, L., Do, T., Ghez, A., et al. 2008, \apj, 688, L17

\bibitem[Meyer {\it et al.}(2009)]{meyer09} Meyer, L., Do, T., Ghez, A., et al. 2009, \apj, 694, L87

\bibitem[Miralda-Escude(2012)]{escude12} Miralda-Escude, J., 2012, submitted to ApJ, eprint arXiv:1202.5496

\bibitem[Miyazaki, Tsutsumi \& Tsuboi (2004)]{Miyazaki04} Miyazaki, A., Tsutsumi, T., Tsuboi, M. 2004, ApJ 611, L97.

\bibitem[Morris(2012)]{morris12} Morris, M., 2012, Nature, 481, 32

\bibitem[Morris, Wang \& Yuan (2011)]{MWY11} Morris, M.R., Wang, Q.D., Yuan, F., eds., 2011, ASP Conf. Ser. Vol. 439, The Galactic Center: A Window to the Nuclear Environment of Disk Galaxies, San Francisco, ASP. 

\bibitem[Moscibrodzka {\it et al.}(2012)]{MSGD12} Moscibrodzka, M., Shiokawa, H., Gammie, C.F., Dolence, J.C. 2012, ApJ 752, L1.

\bibitem[Muno {\it et al.} (2007)]{Muno07iron} Muno, M.P., Baganoff, F.K., Brandt, W.N., Park, S., Morris, M.R. 2007, ApJ 656, L69.

\bibitem[Murakami {\it et al.}(2001a)]{Murakami01a} Murakami, H., Koyama, K., Tsujimoto, M., Maeda, Y., Sakano, M. 2001a, ApJ 550, 297.

\bibitem[Murakami, Koyama \& Maeda(2001b)]{Murakami01b} Murakami, H., Koyama, K., Maeda, Y. 2001, ApJ 558, 687.

\bibitem[Murray-Clay \& Loeb(2012)]{loeb12} Murray-Clay, R. A., Loeb, A., 2012, submitted to Nature, eprint arXiv:1112.4822

\bibitem[Muzi\'c {\it et al.}(2010)]{Muzic:2010fk} Muzi\'c, K., Eckart, A., Sch\"odel, R., Buchholz, R., Zamaninasab, M., Witzel, G. 2010, \aap 521, A13.

\bibitem[Nakajima {\it et al.}(2009)]{Nakajima09} Nakajima, H., Go Tsuru, T., Nobukawa, M., Matsumoto, H., Koyama, K., Murakami, H., Senda, A., Yamauchi, S. 2009, \pasj 61, 233.

\bibitem[Nobukawa {\it et al.}(2011)]{Nobukawa11} Nobukawa, M., Ryu, S.G., Go Tsuru, T., Koyama, K. 2011, ApJL 739, L52.

\bibitem[Nowak {\it et al.}(2012)]{Nowak12} Nowak, M.A., Neilsen, J., Markoff, S.B., Baganoff, F.K. et al. 2012, preprint.

\bibitem[Odaka {\it et al.}(2011)]{Odaka11} Odaka, H., Aharonian, F., Watanabe, S., Tanaka, Y., Khangulyan, D., Takahashi, T. 2011, ApJ 740, 103.

\bibitem[Olling \& Merrifield(2000)]{olling00} Olling, R. P., Merrifield, M. R. 2000, MNRAS, 311, 361.

\bibitem[Park {\it et al.}(2004)]{Park04X} Park, S., Muno, M.P., Baganoff, F.K., Maeda, Y., Morris, M., Howard, C., Bautz, M.W., Garmire, G.P. 2004, ApJ 603, 548.

\bibitem[Perets {\it et al.}(2007)]{Perets07} Perets, H.B., Hopman, C., Alexander, T. 2007, ApJ 656, 709.

\bibitem[Perets {\it et al.}(2009)] {Perets09} Perets, H.B., Gualandris, A., Kupi, G., Merritt, D., Alexander, T. 2009, ApJ 702, 884.

\bibitem[Ponti {\it et al.}(2010)]{Ponti10} Ponti, G., Terrier, R., Goldwurm, A., Belanger, G., Trap, G. 2010, ApJ 714, 732.

\bibitem[Porquet {\it et al.}(2003)]{Porquet03} Porquet, D., Predehl, P., Aschenbach, B., et al. 2003, \aap 407, L17.

\bibitem[Porquet {\it et al.}(2008)]{Porquet08} Porquet, D., Grosso, N., Predehl, P., et al. 2008, \aap 488, 549.

\bibitem[Psaltis(2004)]{psaltis04} Psaltis, D., 2004, In: X-RAY TIMING 2003: Rossi and Beyond. AIP Conference Proceedings, Volume 714, 29.

\bibitem[Quataert(2002)]{Quataert02} Quataert, E. 2002, ApJ 575, 855.

\bibitem[Reid {\it et al.} (2009)]{Reid09SgrB2} Reid, M.J., Menten, K.M., Zheng, X.W., Brunthaler, A., Xu, Y. 2009, ApJ 705, 1548.

\bibitem[Revnivtsev {\it et al.} (2004)]{Revnivtsev04} Revnivtsev, M.G., Churasov, E.M., Sazonov, S. Yu., Sunyaev, R.A., Lutovinov, A.A., Gilfanov, M.R., Vikhlinin, A.A., Shtykovsky, P.E., Pavlinsky, M.N. 2004, A\&A 425, L49.

\bibitem[Rodr\'iguez-Fern\'andez {\it et al.}(2004)]{RodFer04} Rodr\'iguez-Fern\'andez, N.J., Mart\'in-Pintado, J., Fuente, A., Wilson, T.L. 2004, \aap 427, 217.

\bibitem[Rubilar \& Eckart(2001)]{rubilar01} Rubilar, G. F., Eckart,  A.  2001, \aap  374, 95.

\bibitem[Sazonov, Sunyaev \& Revnivtsev(2012)]{SSR12} Sazonov, S., Sunyaev, R., Revnivtsev, M. 2012, \mnras 420, 388.

\bibitem[Schartmann {\it et al.}(2012)]{schartmann12} Schartmann, M., et al. 2012, submitted to ApJ, eprint arXiv:1203.6356

\bibitem[Sch\"odel {\it et al.}(2002)]{schoedel02} Sch\"odel, R.,  et al. 2002, Nature, 419, 694.

\bibitem[Shcherbakov \& Baganoff (2010)]{SB10} Shcherbakov, R.V., Baganoff, F.K. 2010, ApJ 716, 504.

\bibitem[Sunyaev \& Churazov(1998)]{SunyaevChurazov98} Sunyaev, R., Churazov, E. 1998, \mnras 297, 1279.

\bibitem[Sunyaev, Markevitch \& Pavlinsky(1993)]{SMP93} Sunyaev, R.A., Markevitch, M., Pavlinsky, M. 1993, ApJ 407, 606.

\bibitem[Terrier {\it et al.} (2010)]{Terrier10} Terrier, R., Ponti, G., B\'elanger, G., Decourchelle, A., Tatischeff, V., Goldwurm, A., Trap, G., Morris, M.R., Warwick, R. 2010, ApJ 719, 143.

\bibitem[Trap {\it et al.}(2011)]{trap11} Trap, G., Goldwurm, A., Dodds-Eden, K., et al. 2011, \aap 528, A140.

\bibitem[Trippe {\it et al.}(2007)]{trippe07} Trippe, S., Paumard, T., Gillessen, S., et al. 2007, MNRAS, 375, 764.

\bibitem[Tsuboi, Handa \& Ukita (1999)]{Tsuboi99} Tsuboi, M., Handa, T., Ukita, N. 1999, ApJS 120, 1.

\bibitem[Uttley \& McHardy(2005)]{uttley05} Uttley, P., McHardy, I. M., 2005, MNRAS, 363, 586 

\bibitem[Valinia {\it et al.}(2000)]{Valinia2000} Valinia, A., Tatischeff, V., Arnauld, K., Ebisawa, K., Ramaty, R. 2000, ApJ 542, 733.

\bibitem[van der Laan(1966)]{vdL66} van der Laan, H. 1966, Nature 211, 1131.

\bibitem[Wardle(2011)]{wardle11} Wardle, M. 2011, in Morris, M.R., Wang, Q.D., Yuan, F., eds., ASP Conf. Ser. Vol. 439, The Galactic Center: A Window to the Nuclear Environment of Disk Galaxies, San Francisco, ASP, p. 450.  

\bibitem[Weinberg {\it et al.}(2005)]{weinberg05} Weinberg, N. N.,  Milosavljevic, M., Ghez,  A. M. 2005, \apj,  622, 878.

\bibitem[Will(2008)]{will08}  Will, C. M. 2008, \apj, 674, L25.

\bibitem[Witzel {\it et al.}(2012)]{witzel12} Witzel, G., et al. 2012, submitted to \apj

\bibitem[Yelda {\it et al.}(2012)]{yelda12}
Yelda, S., et al. 2012, in preparation

\bibitem[Young(1976)]{young76}
Young, P. J. 1976, \prd, 14, 3281

\bibitem[Yu {\it et al.}(2011)]{Yu11} Yu, Y.-W., Cheng, K.S., Chernyshov, D.O., Dogiel, V.A. 2011, \mnras 411, 2002.  

\bibitem[Yusef-Zadeh, Law \& Wardle (2002)]{YLW02} Yusef-Zadeh, F., Law, C., Wardle, M. 2002, \apjl 568, L121.

\bibitem[Yusef-Zadeh {\it et al.}(2006a)]{Yusef06a} Yusef-Zadeh, F., Bushouse, H., Dowell, C.D., et al. 2006a, ApJ 644, 198.

\bibitem[Yusef-Zadeh {\it et al.}(2006b)]{Yusef06} Yusef-Zadeh, F., Roberts, D., Wardle, M., Heinke, C.O., Bower, G.C. 2006b, ApJ 650, 189.

\bibitem[Yusef-Zadeh {\it et al.}(2007)]{Yusef-Zadeh07a} Yusef-Zadeh, F., Muno, M., Wardle, M., Lis, D.C. 2007, \apj 656, 847.

\bibitem[Yusef-Zadeh {\it et al.}(2008)]{Yusef08} Yusef-Zadeh, F., Wardle, M., Heinke, C., Dowell, C.D., Roberts, D., Baganoff, F.K., Cotton, W. 2008, ApJ 682, 361.  

\bibitem[Yusef-Zadeh {\it et al.}(2009)]{Yusef09} Yusef-Zadeh, F., Bushouse, H., Wardle, M., et al. 2009, ApJ 706, 348.

\bibitem[Yusef-Zadeh {\it et al.}(2012)]{Yusef-Zadeh12} Yusef-Zadeh, F., Wardle, M., Dodds-Eden, K., et al. 2012, AJ 144, 1. 
 
\bibitem[Zamaninasab {\it et al.}(2010)]{zamani10} Zamaninasab, M., et al. 2010, A\&A, 510, A3.

\bibitem[Zhao {\it et al.}(1989)]{Zhao89} Zhao, J.-H., Ekers, R.D., Goss, W.M., Lo, K.Y., Narayan, R. 1989, in IAU136: The Center of the Galaxy, M. Morris, ed., Kluwer: Dordrecht, p. 535.

\bibitem[Zhao {\it et al.}(2003)]{Zhao03} Zhao, J.-H., Young, K.H., Herrnstein, R.M., Ho, P.T.P., Tsutsumi, T., Lo, K.Y., Goss, W.M., Bower, G.C. 2003, ApJ 586, L29.

\bibitem[Zhao {\it et al.} (2009)]{Zhao09} Zhao, J.-H., Morris, M.R., Goss, W.M., An, T. 2009, ApJ 699, 186.  

\bibitem[Zucker {\it et al.}(2006)]{zucker06} Zucker, S., et al.  2006, \apj, 639, L21.

\end{thebibliography}
\end{document}